Opportunities and Challenges for Machine Learning in Materials Science

Dane Morgan and Ryan Jacobs

Department of Materials Science and Engineering
University of Wisconsin-Madison, Madison, WI, 53706
ddmorgan@wisc.edu, rjacobs3@wisc.edu

**Keywords:** machine learning, materials informatics, materials science, model assessment, applicability domain, model errors, materials discovery, materials design, artificial intelligence

**Abstract:** Advances in machine learning have impacted myriad areas of materials science, ranging from the discovery of novel materials to the improvement of molecular simulations, with likely many more important developments to come. Given the rapid changes in this field, it is challenging to understand both the breadth of opportunities as well as best practices for their use. In this review, we address aspects of both problems by providing an overview of the areas where machine learning has recently had significant impact in materials science, and then provide a more detailed discussion on determining the accuracy and domain of applicability of some common types of machine learning models. Finally, we discuss some opportunities and challenges for the materials community to fully utilize the capabilities of machine learning.

# 1 Introduction

Machine learning (ML) is playing an increasing role in our society, and more specifically in materials science and engineering (MS&E). This review seeks to provide a brief introduction to ML and its growing roles in an array of aspects of MS&E, as well as a more detailed discussion of some of the challenges and opportunities associated with using ML for predicting materials properties and accelerating the design of new materials. We hope it will therefore be of value for both the novice and experienced user.

ML can be defined as the use of computer systems that do not require explicit programming to learn about the task they are completing. ML falls into two major categories, unsupervised and supervised learning. Unsupervised ML learns properties of data without any human guidance, for example, putting data into groups (clustering) or finding dominant directions of data variation in high-dimensional space (principal component and linear discriminant analysis). These unsupervised methods have the advantage of being able to analyze data with no need for humans to explicitly label the data, which is often a time- and resource-intensive endeavor. In contrast, supervised ML uses labeled data to learn a relationship between an output $Y$ and an input $X$, and is supervised in the sense that it must be told the values of $Y$ and the corresponding values of $X$. This type of learning includes traditional regression (e.g., multivariate linear regression (MVLR)), as well are more recent methods such as deep learning (to be discussed more later) to find objects in an image. Supervised learning typically requires human input to label the data (e.g., labeling objects in an image) although sometimes the computer can generate labels itself, e.g., from a simulation.

Tools and applications of ML have undergone an extremely rapid growth in approximately the last 20 years, with a series of stunning achievements that have been widely reported, including ML algorithms exhibiting superhuman capability at Chess[1], Go[2], Poker[3,4] and Jeopardy,[5] as well as other computationally demanding tasks like image recognition, autonomous driving, and real-time language translation. Many of these capabilities were until recently thought to be grand challenges likely to remain inaccessible for decades.[6] A detailed discussion of the cause of this transformation is beyond the scope of this paper, but likely involves a confluence of ever increasing computing power (e.g., GPUs), the exploding scale and accessibility of data (e.g., cloud resources), and multiple significant algorithmic advancements with quite general applicability (e.g., deep learning for images and natural language processing). These influences have now become self-reinforcing, with computing, data, and algorithms all taking advantage of, and driving innovations in, the other areas to enable increasingly impressive applications. For instance, in 2016 Google announced it had deployed Tensor Processing Unit (TPU) chips specifically designed to enable fast training of deep neural networks, and demonstrated improved performance by a factor of 15-30 compared to leading GPU technology.[7] Another example is the development of neuromorphic chips (e.g., the True North chip from IBM and the Pohoiki Beach chip from Intel), which intrinsically have a neural network-like functionality and, when compared to standard CPU and GPU chips, can be dramatically more efficient in terms of power consumption.

ML: machine learning
MS&E: materials science and engineering
MVLR: multivariate linear regression
GPU: graphics processing unit
TPU: tensor processing unit
MGI: materials genome initiative
ICME: integrated computational materials engineering
NOMAD: Novel Materials Discovery
MaX: Materials design at the Exascale
MGE: Materials Genome Engineering (MGE)

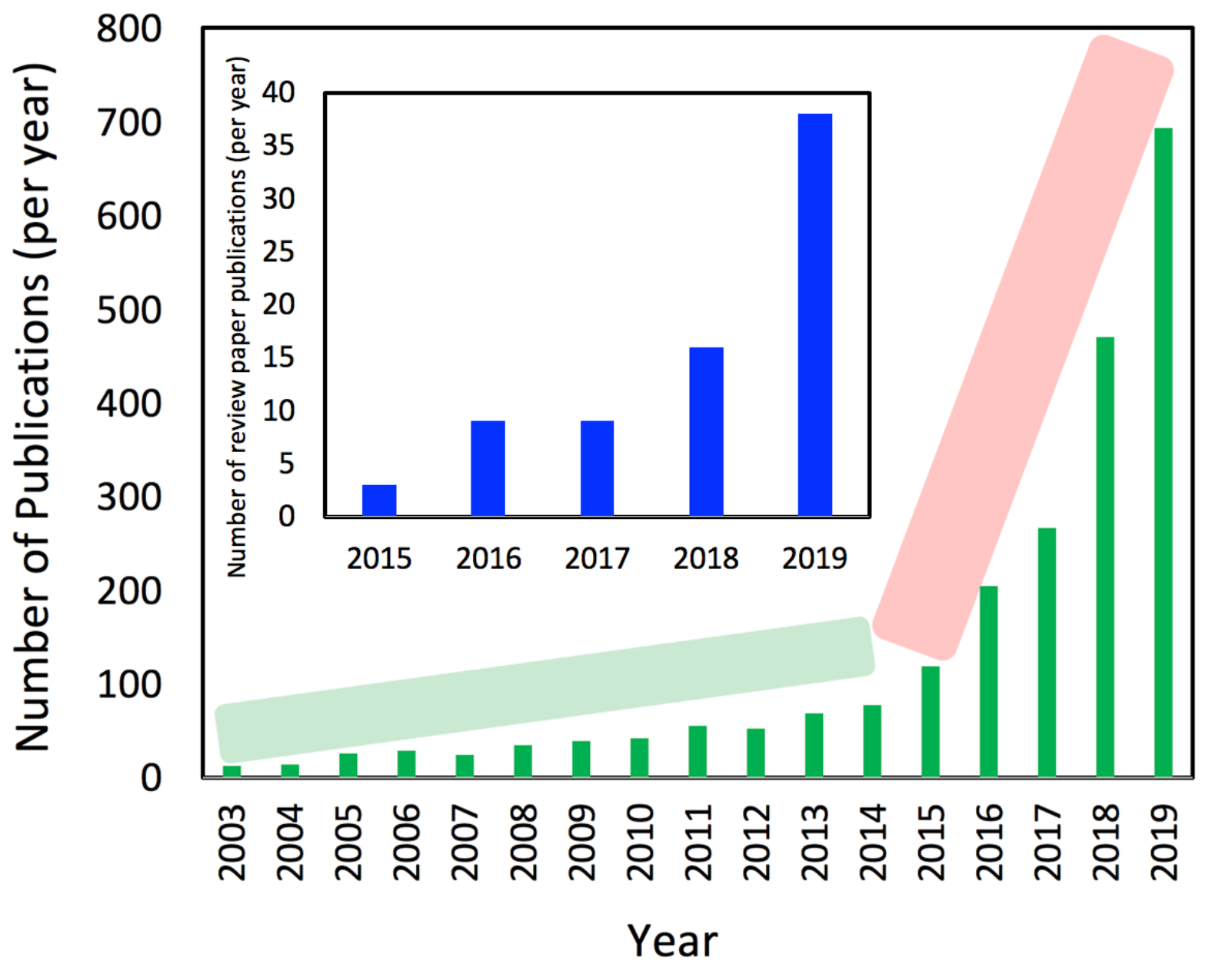

Figure 1. Growth of ML in MS&E as seen in overall and review (inset) publications.

The potential impact of ML, particularly in critical and economically massive areas such as high-tech businesses, manufacturing, national defense and healthcare has led to resources being committed to develop the ML ecosystem (computing, data, algorithms, and software) on the scale of billions of dollars per year in the U.S. and other countries. These resources have created an extraordinary opportunity for MS&E researchers to benefit from this ecosystem with only modest investment, somewhat analogous to the way computational MS&E has been enabled by inexpensive commodity processors developed for other fields. In particular, open source software implementing state-of-the-art ML algorithms is widely available, often developed by leading ML companies (e.g., Google, Facebook), as are relatively inexpensive computing resources, including GPUs for deep learning. These techniques can be integrated with the rapidly increasingly world of materials data, which is being generated by new instruments and simulations as well as being shared through new cloud-based resources. Worldwide growth of frameworks and initiatives such as Integrated Computational Materials Engineering (ICME)[8–10], the Materials Genome Initiative (MGI),[11,12] Novel Materials Discovery (NOMAD), Materials design at the Exascale (MaX), and the Materials Genome Engineering (MGE) program in China have helped support a growing computation and data infrastructure in the MS&E community which is poised to take advantage of the new ML ecosystem.

The renaissance in computing power, data production and dissemination, and ML tools and their availability, is creating very rapid growth in ML in MS&E, particularly since 2014 (Figure 1)[1]. An examination of a logarithmic plot for the data in Figure 1 suggests that since 2014 we have seen exponential growth of the form A(papers)×exp(t/B(years)), suggesting a doubling about every 1.6 years. No single review can cover the broad range of areas and methods being pursued in detail, and in this review we include both a high-level overview of areas and trends and then a more detailed discussion of one specific central concern. In particular, in Sec. 3 we provide a brief discussion of major ML application areas in MS&E to help guide researchers attempting to understand the landscape and perhaps take first steps into a given area. In Sec. 4 we focus on the supervised learning models for property prediction, which is one of the most frequent uses of ML in MS&E, and describe some of the best practices for model development and assessment. Sec. 5 provides guidance on our summaries of useful tools for ML in MS&E. Throughout this review we focus on recent results and present opportunities and challenges, and then in Sec. 6 we offer some more speculative thoughts on longer term future opportunities and challenges. All data associated with this paper that are shared online are described, with appropriate links, in Sec. 7. This includes catalogues of recent review papers and ML software tools shared via Figshare so they can be easily updated in the future. We include in Supporting Information (SI) a summary of useful infrastructure information for ML in MS&E, including a detailed list of more than 70 recent reviews (Sec. S1), software tools for general and MS&E specific ML applications (Sec. S2), journals that frequently publish ML MS&E applications (Sec. S3), introductory discussion of common ML models used in MS&E and recent strategies for working with small datasets (Sec. S4), some useful benchmark tests of model performance for comparison with naïve reference points (Sec. S5), and some more in-depth comparison of model error estimates (Sec. S6). The catalogues of recent review papers and ML

SMILES: simplified molecular-input line-entry system

NN: neural network

KRR: kernel ridge regression

GPR: gaussian process regression

RFDT: random forest decision tree

QSAR: quantitative structure-activity relationships

QSPR: quantitative structure-property relationships software tools are also posted online via Figshare (see link in Data Availability in Sec. 7) so they can be easily updated in the future.

[1] Publications/year in materials informatics (Web of Science search of ("machine learning" or "artificial intelligence" or “materials informatics” or “data science”) and (“materials"), scaled by 0.75 to correct for average rate of errors. Review publications/year from manual citation search in Google Scholar and Web of Science.

# 2 Some notation

In order to avoid repeating notation in multiple locations we will introduce it here and use it consistently in this review. We will frequently consider supervised regression problems where we assume our data has the original form ($X$,$Y$), where $X$ is matrix of features and $Y$ is a vector of target values. Commonly, each row of $X$ corresponds to a system (e.g., a material structure and composition) to be modeled and each element in that row is a value describing some feature of the system (e.g., amount of Cu), and $Y$ is a vector of target properties to be modeled (e.g., band gap). $X$ typically starts in the form of a human-relevant simple description (e.g., just composition and structure, or a simplified molecular-input line-entry system (SMILES) string) and corresponding features in a numerical form must be generated (this process is sometimes called featurization and is discussed in Sec. 4.2). The relationship between $X$ and $Y$ can be written as $Y = F(X) + \epsilon$, where $\epsilon$ is a noise term (with mean zero and variance $\sigma^2$) and we seek to use ML to construct a model for $F(X)$. We write this model as $\hat{F}(X)$ and its predictions as $\hat{Y}$. Given some new $X^*$ one can use the ML model to predict a corresponding target value, $\hat{Y}^* = \hat{F}(X^*)$.

In general, $\hat{F}$ can be specified by its model type, parameters, and hyperparameters. Model type refers to the overall functional forms used, e.g., linear regression or neural networks (NNs). Model parameters are the values that define the specific instantiation of the model and are fit during the training process, e.g., coefficients of linear terms or weights in a NN. Model parameters can generally be fit by some highly efficient method, e.g., matrix inversion for linear models or backpropagation for NNs. Model hyperparameters are similar to model parameters but cannot be easily optimized through an efficient method, and are therefore typically treated separately from the model parameters and searched in a more restricted manner, e.g., with a simple grid search, with full optimization of model parameters for each evaluation of model hyperparameters. Examples of model hyperparameters include number of terms in a polynomial regression or number of layers in a NN.

# 3 Where and how is ML impacting MS&E?

This section provides a high-level summary of some of major areas where ML is being applied in the field of MS&E, and some representative examples from recent studies are showcased in Figure 2.

## 3.1 Property prediction and materials discovery and design

### 3.1.1 Property prediction

One of the most common and easy to understand uses of ML in MS&E is predicting new materials data from existing databases through regressing $Y$ on $X$ followed by prediction of $\hat{Y}^* = \hat{F}(X^*)$ for new data (See Sec. 2 for notation). There is no unique approach to assigning feature vectors in $X$ to represent a material and this is a critical challenge we discuss in detail in Sec. 4.2. This overall approach can be used to extend almost any database to new systems, allowing prediction of new data, rapid exploration of large spaces, and iterative optimization to find new materials (sometimes called active learning). The use of ML in MS&E has been applied to predict myriad materials properties for many classes of materials. A representative but not exhaustive list of recent studies include the prediction of bulk stability of perovskite and garnet oxides and elpasolites,[13–16] formability of novel ternary compounds,[17,18] superconducting critical temperatures of complex oxides,[19,20] melting points of unary and binary solids,[21] dielectric properties of perovskites and polymers,[22,23] formability of novel half- and full-Heusler intermetallic compounds,[24,25] casting size of metallic glass alloys,[26] electronic bandgap of different classes of inorganic materials such as oxides and covalent semiconductors,[27282930] stability and bandgap of halide perovskites for solar cells,[31–33] dilute metal element solute diffusion barriers in an array of metallic hosts,[34,35] electromigration of impurity elements in metals,[36] scintillator materials,[37] and piezoelectric materials with high electrostrains,[38] among others. Figure 2A shows an example heatmap detailing the number of newly discovered ternary oxide materials across chemical space, which predictions were obtained by using ML to inform the probability a ternary oxide will form. [17] When too little data is available for regression, clustering can still provide a tool by grouping similar materials based on their features. To the extent that these groups share properties, such a clustering can provide powerful predictions, and some uses for finding phase diagrams and allotropes are summarized in the review from Ramprasad et al.[39] Some effective and widely-used regression methods used in the materials data studies listed above include MVLR,[36] kernel ridge regression (KRR),[1432] Gaussian process regression (GPR),[2738] ensemble methods such as random forest decision trees (RFDTs) and gradient boosted regression,[243340] and both basic and deep learning neural networks.[34131441] We note here that for readers less familiar with these different ML methods, we have included more introductory discussion of these different model types in Sec. S4 of the Supporting Information, which is also mentioned in Sec. 4.3. In addition, more detailed information on these general ML methods are covered in the references [42–45].

It is worth noting that many present ML approaches for predicting structure-property-performance relationships of materials in MS&E can be viewed as part of, or emerging from, the field of study known as quantitative structure-activity relationships (QSAR) (and the closely related field of quantitative structure-property relationships (QSPR)).[4647] QSAR/QSPR have used data science tools for over 100 years for correlating physical and molecular properties of chemical substances and their associated properties, from biological activity to boiling point, and therefore present well-established best practices and powerful techniques that can provide excellent guidance to the MS&E community.

### 3.1.2 Materials discovery and design

ML has built on its strength in property prediction (Sec. 3.1.1) to enable the discovery, design and development of novel materials spanning an array of applications and materials classes by providing new understanding of key chemical or physical relationships governing properties of interest. As a concrete example, in the field of halide perovskites for solar photovoltaics, the use of ML on data has resulted in assessment of chemical trends (e.g. halogen content and alkali vs. organic species content) on properties such as the bandgap and stability, and resulted in the prediction of new promising halide perovskite materials such as $Cs_2Au^{1+}Au^{3+}I_6$ and $NH_3NH_2InBr_3$, the former of which has been investigated in detail as a promising solar material.[32334840] In addition, materials data predictions from ML on a large space of Br- and Cl-based elpasolite compounds led to the discovery of numerous new promising scintillator materials and also reproduced more than 20 known well-performing scintillators. In this case, insights from ML provided rational material composition changes to realize a favorable placement of the $Ce^{3+}$ 4f and 5d levels within the material bandgap, a necessary design criterion for scintillators.[37]

In some cases ML is used as an integral guide to the data collection effort, e.g. in active learning, where iterative design of experiments (or simulations) is performed using ML property models and carefully tuned optimization approaches.[38,49–5354] More specifically, active learning is a method which seeks to balance exploitation of information contained in existing data in an ML model (i.e. data points with the best predictions) and exploration of less-sampled portions of the design space (i.e. data points likely to have high model uncertainties). Active learning is used to obtain a target outcome as efficiently as possible by first quickly sampling potential regions of interest to construct an initial ML model, followed by adaptive sampling of the exploitation-exploration tradeoff to maximize the expected improvement of the ML model for finding the target, thus optimizing the experimental objective (e.g. finding a new material with highest electronic bandgap) with the fewest number of measurements. Active learning methods have yielded numerous success stories, e.g., a new Pb-free piezoelectric material with the largest measured electrostrain in the $BaTiO_3$ family[38] and new polymers with high glass transition temperatures, the latter result obtained by starting from only remarkably small training dataset of just 5 materials.[55]

An exciting and fairly new area for materials discovery using ML is the integration of autonomous high-throughput experimentation conducted by robots with on-the-fly decision making guided by ML model predictions made using active learning techniques.[56–65] This integration has the potential to perform guided exploration of large materials spaces with limited to no human intervention, greatly accelerating rates of materials discovery as well as potentially supporting work with materials or in environments that are inhospitable to humans[59] and reducing human biases in materials searches.[59,63] These approaches have had some notable recent successes. Duros et al.[63] explored new approaches for the synthesis and crystallization of a new polyoxometalate compound, and demonstrated that the purely machine-based search covered about 6 times larger parameter space to realize crystallization than that explored by humans with a prediction accuracy of whether the compound will crystallize about 5% higher than that obtained by humans. Granda et al.[56] demonstrated an ML-guided organic synthesis robot which was able to predict the outcome of untested chemical reactions with greater than 80% accuracy, and then was able to construct prioritized lists of new reactions to attempt based on their evaluated likelihood to produce the desired products. A further outcome of this work was the identification of unusual reaction mixes, which were later evaluated by human researchers, leading to the discovery of previously unknown chemical reactions. Finally, Nikolaev et al.[60] developed a robot scientist named the Autonomous Research System (ARES) that specialized in the autonomous growth and characterization of carbon nanotubes, a model problem due to its complex coupling of synthesis and processing to resulting structure-property relationships (e.g., example nanotube diameter, helicity and the effects of these parameters on the nanotube electronic properties). Figure 2B contains a photograph showing the lab setup of the ARES instrument. ARES successfully optimized nanotube synthesis in a high dimensional design space and determined the correct parameters to maintain accurate growth rate control, thus demonstrating the potential utility and possible disruptive potential of ML-guided robot scientists in MS&E.

GAN: generative adversarial network

A particularly interesting area is developing new materials with generative models such as variational autoencoders and generative adversarial networks (GANs). These methods are particularly well-suited to execute the paradigm of inverse materials design, where the desired material characteristics are first enumerated and candidate materials are suggested and evaluated on-the-fly.[66–69] Inverse design creates the challenge of the exploration of an exceedingly large chemical space, which can be partly overcome by the use of GANs to automatically suggest and evaluate novel molecules and materials for a desired application.[59] Concrete successes have already been demonstrated in this area, for example the CrystalGAN[70] model which was used to generate, screen and subsequently discover new stable hydride compounds for solid-state hydrogen storage applications. The objective-reinforced generative adversarial network for inverse-design chemistry (ORGANIC) model was shown to be successful in predicting new high melting point organic molecules.[71] Figure 2C shows a schematic of the ORGANIC model, which consists of separate generator (discriminator) neural networks used to suggest new molecular structures (predict desired molecular properties), respectively, where the generation of new candidate molecules is informed by the discriminator and the reinforcement algorithm. Finally, the Reinforced Adversarial Neural Computer (RANC) was found to outperform ORGANIC and function as a valuable tool for discovery of novel molecules for drug design and development.[72]

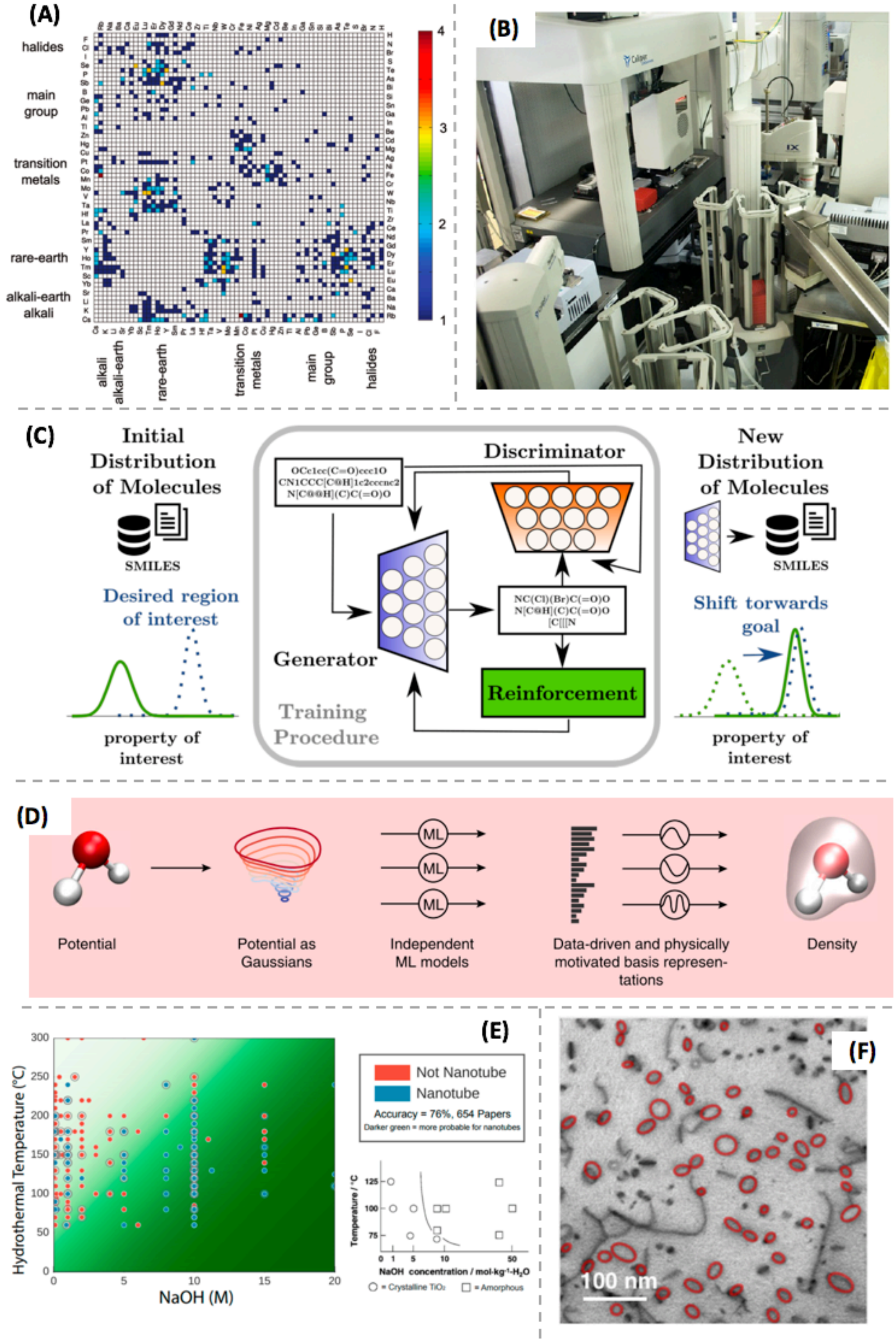

Figure 2. (A) Heatmap showing number of newly discovered ternary oxide materials across chemical space, (reprinted with permission from Ref. [17] copyright 2010 American Chemical Society). (B) photograph of an autonomous synthesis and characterization robot, (adapted from Ref. [65]). (C) Overview of usage of the ORGANIC GAN model for molecular design (adapted from Ref. [71]). (D) Scheme to represent the Hohenberg-Kohn map using machine learning models (adapted from Ref. [127]). (E) Machine-learned probability of nanotube synthesis (green shades) compared with experimental outcomes from text mining (reprinted with permission from Ref. [97] copyright 2017 American Chemical Society. (F) Model-labeled elliptical irradiation defects in a characterized steel micrograph (adapted from Ref. [75]).

## 3.2 Materials characterization

Materials characterization tools are increasingly producing data on scales of quantity and complexity which outstrips human ability to manage and interpret, and ML methods are being used to process and analyze this data. For example, Voyles recently reviewed a number of applications in electron microscopy,[73] pointing out uses of ML in image improvement (e.g., denoising, drift and distortion correction) and analysis (e.g., spectral demixing and clustering to identify features). A number of studies have recently applied deep learning machine vision techniques to electron microscopy images, e.g., to cluster materials based on microstructure images (see references in [74]) and to identify defects in images,[75,76] in multiple cases with apparently human levels of accuracy. For example, Figure 2F shows dislocation loops in electron micrographs of a steel alloy identified by a deep learning model, the accuracy of which was as good or better than that of domain-specific expert humans.[75] ML has also been applied to X-ray diffraction data, e.g., using deep learning to accurately perform identification of space-group, extinction-group and crystal-system from X-ray powder diffraction patterns.[77] Other intriguing examples have shown how machine learning could replace more challenging measurements or calculations with simpler ones. As an experimental example Stein, et al. demonstrated that a variable auto-encoding approach could quite accurately reproduce UV-vis spectra from simple images of a thin film generated with a commercial scanner.[78] In simulation, Combs, et al. recently demonstrated that a MVLR model could correlate low and high fidelity scanning tunneling electron microscopy (STEM) image modeling, allowing approximations to full multislice simulations of nanoparticles millions of times faster than a full STEM image simulation. ML appears likely to provide many paths toward accelerated characterization through simplified experiments and computations, and automated analysis, reducing time spent in traditional methods and enabling processing of the enormously large data streams coming from newer and next-generation characterization instruments.

> STEM: scanning tunneling electron microscopy
>
> NLP: natural language processing

## 3.3 Knowledge extraction via text mining

Natural language processing (NLP) tools are central to text and speech extraction and recognition, enabling AI-related speech tools like Apple’s Siri and Amazon’s Alexa and real-time language translation. Numerous open source NLP tools currently exist, for example the word2vec[79] and Global Vector (GloVe)[80] packages, and tools to conduct sentiment analysis using deep convolutional neural networks.[81] NLP, text extraction and sentiment analysis (i.e. the characterization of subjective information such as opinions, communicated through text) have seen widespread use, for instance in computational biology and biomedical research,[82,83] genetics,[84] healthcare,[85] and social science,[86] but work has been much more limited in materials.

A basic NLP analysis in MS&E can be considered in three steps. First, one maps words to real-valued vectors, a process called embedding, which can be done with unsupervised learning and requires significant time and large data sets. However, once completed, such embeddings can be reused in many applications, and multiple MS&E specific embeddings are already available.[87–89] Given an embedding, the second step is to train a NLP model to recognize target information using embeddings, typically with supervised training on a set of expert-annotated sentences, some of which are now being made open-source to encourage democratization of the NLP ML model training process.[90] The second step treats the sentence as a sequence of words, which are converted to a sequence of vectors, and the model is trained to predict the correct annotated categories (e.g., identify text "Fe" as category "metal") on the training data. Models are typically recursive, convolutions, or transformer NNs.[91–93] A common third step then applies grammar rules to understand connections between identified words in a dependency parse tree, e.g., allowing one to determine if a given value is describing a given property in a sentence.[94]

Software packages like ChemDataExtractor[95] are being developed to enhance standard NLP approaches with the ability to parse unstructured text in complex scientific publications, for example chemical formulas or domain-specific words or abbreviations (e.g. the meaning of the occurrence "UV-Vis" spectroscopy). One area where text mining is playing an increasingly large role in MS&E is synthesis.[96] Recent text extraction studies have resulted in useful guidance of key experimental parameters needed for optimal materials synthesis, for example in creating $TiO_2$ nanotubes[97] as shown in Figure 2E, synthesis of new perovskite materials,[98] and aggregated synthesis parameters for 30 different oxide materials systems,[89]. These studies have also provided insights on best writing practices to facilitate efficient transfer of knowledge which is machine-readable,[99] and understanding of best synthesis practices through graph representations.[100] As opposed to the above examples which conducted NLP using supervised methods with annotated studies as training data, Tshitoyan et al.[87] demonstrated the successful use of unsupervised techniques in extracting structure-property and chemical relationship information, and demonstrated these tools can aid in future materials discovery by codifying knowledge contained in past publications. Another impressive example is the recent book *Lithium-Ion Batteries: A Machine-Generated Summary of Current Research*, a review extracted from over 150 papers on Li-ion batteries which was generated by a machine learning model.[101] This work suggests a future where information aggregation in topical reviews could be automatically delivered in a very human understandable form, significantly accelerating the process of learning new areas.

## 3.4 Machine Learning for molecular simulation

### 3.4.1 Interatomic potentials

DFT: density functional theory

MLP: machine learning potential

AIMD: ab initio molecular dynamics

CNN: convolutional neural network

Atomistic scale simulations of molecules and condensed phases typically find the interaction between classically-treated nuclei through Hamiltonians that are either based on approximate solutions to the Schrödinger equation for electrons or that coarse-grain quantum electronic effects into an effective interatomic potential. Interatomic potentials are typically about $10^3$-$10^6$ times faster than common quantum methods (e.g., density functional theory (DFT)) but finding and parametrizing appropriate functional forms to treat systems with complex electronic behavior (e.g., with charge transfer, bond breaking, multiple types of hybridization, etc.) is very challenging. Replacing interatomic potential functional forms and fitting procedures with those from ML offers the alluring possibility of both greatly reducing the time and expertise required for developing potentials and perhaps enhancing their accuracy. In the past decade, many researchers have used ML for generation of interatomic potentials (referred to as a "machine learning potentials" (MLPs)) which have enabled studies of larger size and longer time than accessible with direct DFT.[102–107108] Generating an MLP is fundamentally a complex regression problem to map the potential energy surface and/or its derivative, the force field, by fitting an ML model to a large training database, typically containing thousands of DFT calculations (often derived from individual time steps of Ab Initio Molecular Dynamics (AIMD) simulations).[103,109–112] Constructing input features for the MLP model (sometimes called atomic structure descriptors or “fingerprints”) is critical and has received a lot of attention over the past decade (see Sec 4.2).[39,105,113,114] As a concrete example of the success of these methods, Botu and Ramprasad trained a KRR model to demonstrate a largescale acceleration of AIMD calculations for bulk and surface slabs of Al.[103] More recently, Bartok et al. have shown using GPR that a MLP can accurately capture the energetics of Si surface reconstructions.[109] Finally, Artrith et al. demonstrated that modeling systems with up to 11 elemental components with neural networks is not only computational feasible buy highly accurate.[110,111] Despite these and many other notable successes, challenges remain, such as the difficulty in obtaining enough high quality DFT data to fit a MLP for complex phenomena (such as grain boundaries, surfaces, cluster defects, or other extended defects) and multiple alloying elements. An additional challenge of using MLPs is similar to that encountered with the construction of empirical potentials: namely how to assess the chemical and physical applicability domain of the MLP and understanding when the MLP may fail.[102,110,115,116] Finally, MLPs are often quite slow compared to many traditional interatomic potentials (e.g. about 1 to 2 orders of magnitude slower),[113] so approaches that can accelerate their evaluation would broaden then utility.

### 3.4.2 Improving and accelerating *ab initio* simulations

*Ab initio* methods (e.g. DFT and hybrid functionals) use approximate solutions to the quantum mechanical equations of electrons to model materials systems and have become some of the most widely used tools in materials and chemical science (DFT is today used in at least 30,000 new research publications every year).[117] However, these methods suffer from limitations of accuracy and speed that significantly inhibit their use, and there have been multiple strategies to apply ML to improve and accelerate the calculation of *ab initio* functionals, each showing significant notable advances. One strategy has focused on improving the accuracy of DFT methods. For example, the work of Nagai et al.[118] used a neutral network to numerically calculate the Hartree exchange-correlation functional in an effort to improve its accuracy, and Bogojeski et al. found that one can efficiently learn the energy differences from DFT and coupled cluster simulations, and use ML to provide a promising avenue to have coupled-cluster-level accuracy and DFT-level speed for physical situations where standard DFT is inadequate.[119] Another strategy is to use ML to learn the computationally expensive portions of solving the Kohn-Sham equations in a DFT calculation, namely contributions to the exchange-correlation energy.[120–122] To this end, Snyder et al.[120] modeled the kinetic energy of a one-dimensional system of non-interacting electrons, which analysis was then extended to more general cases,[121] and Mills et al. showed that a convolutional neural network (CNN) can learn the mapping between the potential energy landscape and the resulting one-electron ground state and kinetic energies.[123124]

The second strategy is to directly learn the electron charge density itself. This strategy has the advantage that it allows one to completely bypass solving the Kohn-Sham equations, and instead rely on the Hohenberg-Kohn theorems which allow one to obtain the total energy (and other properties) directly from the charge density.[125–129] Figure 2D shows a representation of using ML and charge densities to bypass the calculation of the Kohn-Sham equations. There have been diverse methods employed to learn the charge density directly. For example, Kajita et al. proposed a method of descriptor generation based on a 3D voxel representation of the electron density for use in CNNs.[126] In contrast, Brockherde et al.[127] and Bogojeski et al.[128] formulated KRR models to directly learn the charge density using a suite of training data, and a final example from Sinitskiy and Pande showed that CNNs trained on low-fidelity charge density data can learn meaningful characteristics of the charge density for a variety of organic molecule chemical environments, enabling predictions with DFT-level accuracy but orders of magnitude faster.[129] Once sufficiently mature, these methods may fundamentally alter the way in which researchers conduct *ab initio* calculations, wherein ML is fundamentally providing quantum mechanical knowledge of complex systems without needing to solve the Schrödinger equation.

# 4 Some challenges and best practices for ML in MS&E

In this section we discuss issues that occur in many ML modeling projects, focusing on supervised regression learning models for property prediction, although many of the issues are similar in other applications. The key steps in a ML workflow broadly include: (1) data collection and cleaning, (2) feature generation and selection (featurization or feature engineering), (3) model type selection, fitting, and hyperparameter optimization, (4) model uncertainty assessment (e.g. performance on test data) and domain applicability, (5) final model predictions. The ML workflow has been discussed extensively in other reviews and a detailed discussion across all parts will not be included here.[39,130] However, we do wish to discuss some critical aspects associated with steps (2)-(4) that we feel are valuable to help the community move toward best practices in ML modeling.

> RMSE: root mean squared error
>
> MAE: mean absolute error
>
> MAPE: mean absolute percentage error
>
> AARD: average absolute relative error
>
> $R^2$: coefficient of linear dependence

## 4.1 Basic Statistics of Accuracy

In many steps of supervised regression learning, ML models are assessed by some statistic related to the differences between the predicted data $\hat{Y}$ and true data $Y$. The equations for these statistics are widely available and will not be given here, but we briefly discuss their effective use. The root-mean squared error (RMSE) is a commonly used error metric, and frequently the error metric that the ML model seeks to minimize. Mean absolute error (MAE) is also useful to calculate and will typically trend with RMSE, but is less sensitive to large errors from outlier predictions and is not smoothly differentiable like RMSE, making it harder to use in some optimizations. Mean absolute percentage error (MAPE) (often called by many different names, e.g., Average Absolute Relative Error (AARD)) is just the absolute error as a percentage of the data point and is also very helpful as the importance of an error is often related to the size the quantity being predicted. It is also important to give RMSE errors relative to the standard deviation of the data set, which is sometimes called the reduced RMSE, as this provides a reasonable representation of the scale of the ML errors with respect to which RMSE should be measured. In particular, the reduced RMSE value for a well-performing ML model should be significantly less than 1 as simply guessing the mean of the predicted data (typically not useful) would yield a reduced RMSE equal to 1.

Another widely used metric is the coefficient of dependence $R^2$, which gives the fraction of variance in the true value that is predictable from the predicted values (the parity plot, which shows predicted vs. actual data, is a very useful plot and gives a graphical feel for $R^2$). $R^2$ is less than or equal to 1, with 1 representing perfect prediction, and can be less than zero for predictions that trend with the opposite sign slope as the true values ($R^2$ technically has no lower bound). Reduced $R^2$ (sometimes referred to as adjusted $R^2$) is given as $R_{red}^2 = 1 - [(1 - R^2)(n - 1)/(n - k - 1)]$ where $n$ is the number of observations and $k$ is the number of features. $R_{red}^2$ is less than or equal to $R^2$ and, since it adjusts for the complexity in the model, it will decrease when terms that have no predictive ability are added. $R^2$ gives a useful overall assessment of model quality, and generally values > 0.7 are desired for a useful model. However, $R^2$ can be misleading, e.g. a few widely separated regions that are fit on average can give a high $R^2$ even when no predictive ability within each region is given by the model.

Overall, we suggest determining at least RMSE, reduced RMSE, MAE, MAPE, $R^2$ and $R^2_{red}$ and generating a parity plot as standard practice, and using the metrics most relevant for your application. RMSE is typically used for choosing the best features and models during ML model development. The exact method of choosing data for fitting and assessing a model with RMSE (or any metric) can be complicated and is described in Sec. 4.4.1.

## 4.2 Feature Engineering

Feature engineering is a key component of developing useful supervised ML models. Features must be machine readable (i.e., vectors of numbers), practical to obtain for the desired application (e.g., they should certainly be significantly easier to obtain than the target property values), capture as much of the relevant variables controlling behavior as possible, and ideally contain limited additional information that is not useful and which may lead to overfitting data and poor predictions. Generally, feature engineering consists of two steps: feature generation and feature selection, each of which will be described here.

A common set of minimal descriptors may include composition and processing conditions (e.g., precursors, annealing temperature or gas pressure), as these can completely specify the final material, although perhaps rather indirectly. Additional characterization information can also be included, e.g., infrared or X-ray diffraction spectral data. While composition specified by weight or atomic percent is useful, it cannot be used to extrapolate to any new elements, since the model will have no knowledge of how to predict effects of that element if it has not appeared in the training database. One solution to this limitation is to represent each element with a feature vector of elemental properties, e.g., melting temperature or electronegativity. These can then be used to generate features for alloys by taking arithmetic- or composition-averaged combinations of the constituent element features, for example constructing the composition-averaged melting point of the elements in a compound. This approach has been codified by the Materials Agnostic Platform for Informatics and Exploration (MAGPIE),[131] which gives a canonical set of elemental properties and arithmetic operations that have proven successful in predicting stable compounds,[14] glass forming ability,[26] and diffusion coefficient, to name a few.[34,35]

For cases where some level of atomic structure (by which we mean atom position and element type) information can be readily determined (e.g., in atomistic modeling or organic molecule descriptions) the atomic structure forms a powerful feature set, as it is likely to play a large or even totally controlling role in setting a property of a molecule or crystal. Direct use of the atomic coordinate vectors and atom types as a feature is inadvisable as it does not satisfy the translational, rotational, and permutation (swapping atoms of same types) symmetries of the system under study, and thus would need a very large amount of data to be trained well enough to reflect these basic symmetries. An array of different feature generation methods have therefore been developed which do satisfy these symmetry requirements. For molecules, these include properties like bond lengths, connectivity, and functional groups, and can include relative atomic position and electronic structure data computed with quantum mechanical atomistic simulations. Such properties have been widely used in QSAR/QSPR analysis. Thousands of basic

MAGPIE: materials agnostic platform for informatics and exploration

ACSF: atom-centered symmetry function

SOAP: smooth overlap of atomic orbitals

BoB: bag of bonds

BAML: bonds, angles, and machine learning

MBTR: many-body tensor representation

QSAR/QSPR features are now available and can be extracted automatically from basic molecular formulae (e.g., SMILES strings) (see Ref. [132] for a summary of recent automated tools for QSAR/QSPR). A number of tools have also been developed for extended systems (i.e., not just molecules) in the context of constructing ML-based potentials (see Sec. 3.4). These features have been validated for particular materials systems and benchmarked against key standard databases (e.g. the QM9 molecule dataset), including, but not limited to: atom-centered symmetry functions (ACSFs),[133] the smooth overlap of atomic orbitals (SOAP) method,[134] partial radial distribution functions,[135] bag of bonds (BoB),[136] bonds, angles and machine learning (BAML),[137] and the many-body tensor representation (MBTR).[138] The streamlined production of many of these features has been implemented in the matminer code package (see Sec. S2 of the SI or the Figshare link in Sec. 7).[139] Another approach to feature generation is graph-based deep learning methods, which first map atomic structure onto a vector of atom descriptors (e.g., type and simple properties, like formal charge) and bond distances and connectivity (the graph), and then merge those descriptions with weighted averaging to ensure flexible joining of the atomic descriptions with the correct bonds. [140–143] These methods work from very basic information and replace the step of invoking human intuition and analysis to generate features with a more automated deep learning generation of a feature map. Finally, we note that one can work from unsymmetrized data if the method itself performs the symmetrization. For example, Nie et al. recently generalized kernel regression approaches to include permutation symmetry and showed it could generate effective energy fitting directly from atomic pair distances.[144]

Once a set of features to represent a dataset have been generated, it is common to select a representative set of features that is large enough to result in low model errors and avoid model underfitting, yet not so large as to incur penalties to overall model accuracy and extrapolative ability due to overfitting. Certain ML models such as polynomial and kernel ridge regression can easily become confused or overfit if too many features are used, but other models like random forest methods (see Sec. 4.3 and Sec. S4 of the SI for more details) intrinsically function as a form of feature selector, as more important features carry heavier weights in the final ensemble of trees compared to less pertinent features. A simple approach is to enumerate all possible feature subsets and select the one minimizing some model error score (e.g., RMSE of a particular cross-validation routine, see Sec. 4.4). For testing up to $M$ features out of $N$ possible features this approach requires $N$ choose $M$ model score evaluations, which is computationally prohibitive for large feature sets. Similar spirited approaches iteratively test one descriptor at a time and then add it to a growing list (forward feature selection) or removes it from a shrinking list (reverse feature selection) based on it resulting in the greatest reduction (or least increase) in the model error score. Forward (reverse) feature selection methods take $N!/(N-M)!$ model score evaluations to find $M$ (remove $M$) features, which is generally tractable for models that are computationally fast to evaluate.

Feature selection usually benefits from a consideration of the physical reasonableness of the features, and features that make no physical sense are obviously a concern, e.g., cost of elements correlating with band gap. Such correlations are likely created by the feature correlating with some other more physical feature(s), but the model not having enough data to select the correct features. Better models can generally be generated by intentionally replacing such features with physically-motivated hand-picked features that perform equivalently well (or better) as automatically selecting features (note that forward (reverse) feature selection are not global optimization methods and can miss optimal feature sets). For example, Liu et al.[36] and Lu et al.[35] found that starting forward selection with an initial physically meaningful feature chosen by human intuition (and known physics) resulted in improved model performance compared to using purely automated forward selection. One can consider iterative exploration of two or more features for addition/subtraction from the feature list, although we are not aware of any examples where this yielded significantly better results and it greatly increases computational cost. In addition to these feature selection methods, other popular dimensional reduction methods take linear combinations of the features to best explain their behavior with fewer variables, often called latent variables (e.g., principal component analysis (PCA), linear discriminant analysis, and factor analysis). Including just the most important latent variables generated by these methods can improve some fits, although the interpretation of the latent variables can be difficult.

PCA: principal component analysis
IRNet: individual residual network
MEGNet: materials graph network
CGCNN: crystal graph convolutional neural network

An important trend to be aware of in ML is the used of deep learning to obtain better results from features without extensive human guidance in both feature construction and/or feature selection (see Sec. 4.3 and Sec. S4 of the SI for more details). Deep learning methods can effectively generate their own feature set (generally called a feature map), often doing so starting from an initially large and rather unstructured set of features (e.g., a vector of pixel intensities or graph-based matrix of atom and bond properties) that are not effective with traditional ML methods. The comparison between more human-crafted vs. machine-learned features in machine vision has largely established the latter as superior in that field, leading to a revolution in the accuracy of machine vision.[145146] While the outcome of this comparison in MS&E problems is not yet clear, there is increasing evidence that deep learning will provide significant improvements. For example, the deep convolutional individual residual network (IRNet)[41] was shown to achieve better performance from a long list of initial features than traditional methods such as RFDTs and ridge regression. Some graph-based deep learning methods, which build feature maps from a very basic initial feature list, have shown comparable or better performance in organic molecule studies than human-crafted traditional features in QSAR/QSPR comparisons, e.g., Message Passing NN frameworks.[147] Similarly, for inorganic materials, the graph-based MatErials Graph Network (MEGNet)[142], SchNet[148] (and SchNetPack[149]) and crystal-graph convolutional neural network (CGCNN)[150,151] have shown performance comparable or better than non-deep learning approaches. Given the success of deep learning in machine vision and language translation, and its already impressive performance compared to more human-crafted features used in traditional methods after just a few years, it seems likely that deep learning-based feature maps will play a major if not dominant role in the future of feature development in ML for MS&E.

## 4.3 Types of Machine Learning Models

CV: cross validation
LO: leave out

The large number of ML models and their many technical details are well-covered in many texts and reviews[42–44,152] and would require more space than available here, so we will not attempt any type of general review of ML models. However, in Sec. S4 of the Supporting Information we provide a short discussion of some of the most commonly used models (with a focus on tools for supervised regression) in MS&E with a goal of highlighting the most salient features for an MS&E researcher.

## 4.4 Model Development and Model Assessment

ML modeling typically has two closely connected but distinct major stages: model development and model assessment. In model development (Sec. 4.4.1) we determine model type, parameters, hyperparameters, and features (see Sec. 2 for definitions of these terms). In model assessment we determine the accuracy of the model for expected use cases, which typically includes assessing the model performance with sampling methods such as cross validation (Sec. 4.4.2), understanding the domain of applicability where the model is expected to be accurate and quantifying error bars in model predicted values to understand expected model uncertainties (Sec. 4.4.3). To help illustrate these important concepts of model applicability domain and assessing model errors more concretely, we provide and discuss an in-depth practical example using ML models trained on data of calculated migration energies for solute elements in metallic hosts (Sec. 4.4.4).

### 4.4.1 Best practices for managing data in model development and assessment

The same model scoring approaches are often used in both model development and model assessment, which can lead to overfitting and overestimation of the model accuracy if one is not careful. This danger is increased when model development involves many degrees of freedom (e.g., many hyperparameters) and where there is limited data to constrain those degrees of freedom. The simple rule to avoid model assessment errors from overfitting is that any data used for model development should not be used for model assessment. To understand how to apply this rule practically it is useful to define three types of data points (where a data point here means a vector of corresponding features $X_i$ and target property value(s) $Y_i$):

- Training data: Data used to determine the optimal model parameters for a given model type, hyperparameters, and feature set.
- Validation data: Data not used in training, which is instead used to assess the error in the model with optimal model parameters determined from fitting the training data. This error is frequently used determine the optimal model type, hyperparameters, and feature set.
- Testing data: Completely left out data not used in training or validation, which is instead used to assess the error in the final optimized model.

First consider the process of model development. We start by dividing the data into training, validation, and test data in some way (we will discuss how to do this most effectively in the practical example in Sec. 4.4.4). A basic fit of the model, with fixed model type, hyperparameters, and feature set, uses training data to find the optimal model parameters that give the lowest possible value of some scoring metric (typically measured with RMSE, so we will use that here) on the training data. This obtained training data RMSE shows how well the model fits the training data, but this error is usually not a good estimate of how the model will fit data outside the training data. This limitation arises because the model often adjusts its many degrees of freedom to properties of the training data that cannot be correctly represented by the model (a process called overfitting), either due to limitations of the model form and features or noise in the data that cannot be modeled. To obtain a reasonable estimate of the model errors on non-training data we can look at how well the model predicts the validation data, for example the validation data RMSE. We can now optimize the model type, hyperparameters, and feature set to minimize the validation data RMSE. The optimal model type, hyperparameters, and feature set can then be used to refit the parameters of this model to the combined training and validation data to get the best possible fitted model without using the test data. The use of any information in model development from the test data, or more generally from a source that would not be available in a corresponding manner during model use, is sometimes called "data leakage", and can lead to overestimating the quality of your model.

### 4.4.2 Model development and assessment with cross validation

Perhaps the most common way to split data into sets for model development and assessment, typically called training and validation sets (defined in Sec. 4.4.1), is the method called cross validation (CV). Splits can be done in many ways, and common approaches include leaving out (LO) one data point or some randomly chosen X% percent fraction (typically called LO one CV or leave out X% CV, respectively), splitting the whole data set into *k* separate equal-sized groups called folds, and iteratively leaving out each fold once (*k*-fold CV), leaving out targeted groups with certain characteristics (LO group CV, sometimes called LO class CV), e.g., all data with a specific chemical composition, and time-split cross validation,[153] which leaves out select data based on the time of their inclusion in the dataset. For LO X% CV and *k*-fold CV one typically chooses which data is in each fold randomly, and this can be done multiple times with different random permutations to ensure good sampling. As discussed in Sec. 4.4.1, The errors in prediction for validation data from models trained on training data, which we will call CV errors, is typically a much better way to assess a model than errors in the training data predictions, as the latter typically show overfitting. CV errors are a common method of model assessment and can be used to develop a model (e.g., RMSE for all folds in 5-fold CV is a common scoring metric used in feature selection, as discussed in Sec. 4.1) and estimate its predictive error, as discussed in more detail in Sec. 4.4.3.

Once an optimized model has been developed, we would like to assess the errors and domain of this optimized model. This model error and domain assessment ideally should not be done with CV scores already obtained using the validation data since these CV scores can be subject to overfitting based on the optimization done in model development. Thus, we need to consider yet another left out data set, the test data, to quantify the model error. Specifically, we take the optimal model type, hyperparameters, and feature set obtained from optimizing the validation data RMSE, and refit the parameters of this model to the combined training and validation data to get the best possible fitted model, and then predict the test data to get the test data RMSE. Because the test data has not been used in any step of the optimization process, the test data RMSE is a good quantification of errors in the final model.

The above approach is often not practical as it is difficult to simply separate out test data and never look at it until the model is finalized. In addition, use of just one training, validation, and test data set may introduce large biases associated with the specific data that ends up in those splits, leading to suboptimal models and error estimates, particularly for smaller data sets. These problems can be avoided by effectively simulating the above steps multiple times with different splits in a method called nested CV. First, you must settle on at least a general model development approach, which includes the types of models you will consider, hyperparameters you will optimize for each model, and features you will explore. Then, you perform CV on all the data, considering each excluded set as test data (level 1 CV), and an additional nested CV (level 2 CV) on the included training and validation data to determine the best model. Each level 1 left out test set then can be considered a true test set in the sense that it was not used in any part of the model development. Many authors effectively perform a level 1 CV just once with ~10-20% of the data left out at level 1 and perhaps also level 2, as multiple folds at level 1 and level 2 can lead to a lot of computation. If the splits are done many times, e.g., with 5-fold CVs for level 1 and 2, it provides a strong sampling across all the data, which is recommended for smaller data sets where it may also be most practical. The nested CV approach is typically not totally rigorous, as researchers will almost inevitably modify aspects of their approach in light of the final results, thereby introducing some level of data leakage and potential overfitting, but nested CV is a practical approximation for quantifying model errors that avoids most effects of overfitting.

One subtlety with the nested CV approach is that while using the level 2 CV to optimize model type, hyperparameters, and feature set one is potentially overfitting to the level 2 CV score with multiple variables, which could lead to some of them actually being incorrectly optimized for truly best performance. A practical example if this, documented in Ref. [154], is that if you optimize hyperparameters for two different model types and then choose between them based on the level 2 CV score, you might choose the less optimal model type simply because it is more overfit. Ideally, one would use many nesting levels and optimize just one property at each level, but this can quickly become impractical, and one nesting level is all that is typically used. Such nesting should be adapted to best meet the specific optimization and assessment needs of your problem.

### 4.4.3 Model domain of applicability and assessing uncertainties in model predictions

Perhaps the most important question one can ask of an ML model is "how accurate is the model for the potential applications I have in mind?" Answering this important question typically has two coupled components, which are (1) an estimate of the domain where the model can be accurately used and (2) an estimate of the uncertainty in the model predicted values (e.g., a standard deviation in prediction accuracy). Regarding (1), the model domain of applicability is a region of feature space outside of which we simply cannot reliably use the model (e.g. using a model trained only on yield strengths of metal alloys to predict yield strengths of polymers). Regarding (2), error estimates provide some form of uncertainty quantification on each value predicted by the model, thus providing more information on the uncertainty of a prediction than simply using the average predicted RMSE of the model from, e.g., a 5-fold CV. In this section, we provide a general introduction to understanding model errors. In Sec. 4.4.4, we illustrate how one may assess the errors and applicability domain of real ML models using GPR and RFDT models fit to a computed database of DFT-calculated dilute impurity diffusion activation energies in a range of metal hosts.

There is not an exact definition of the domain of applicability of a model. We propose that a useful definition which captures what is often desired in defining a domain is the set of data points for which uncertainty can be quantified at a desired level (e.g., that the standard deviation is known within 20%). One might intuitively want to determine a domain of applicability based on some criterion of maximum acceptable errors. However, such screening is only possible if the errors are accurately known, so it is necessary to know the domain in the sense defined above before applying any further constraints on desired error magnitudes. There are many methods to assess domain of applicability based on some measure of distance of the features of a potential data point from those in the model training data, e.g. within the convex hull of the feature space (a number of methods are summarized in Ref. [155]). However, these methods all rely on distance metrics of uncertain validity for the specific problem being studied and require somewhat arbitrary cutoffs, and so are difficult to apply for more than a qualitative guide on where you might consider the model to be at risk of being not applicable. We believe some combination of distances in feature space from training data and predicted error values are likely to provide the best guidance on domain and error estimates. However, predicted error values are more immediately and obviously useful for assessing models, and therefore here we discuss in more detail common methods used to establish some type of error bar on the predictions and their use in establishing model domain, each of which has certain strengths and limitations.

To better understand model prediction errors, it is useful to start with the well-known bias-variance-noise decomposition of the error. Following the definitions in Sec. 2 one can rigorously decompose the expected squared error for prediction on a new point $X^*$ as

$$E\left[\left(F(X)+\epsilon-\hat{F}(X)\right)^2\right]=\left(E\left[\hat{F}(X)\right]-F(X)\right)^2+E\left[\left(\hat{F}(X)-E\left[\hat{F}(X)\right]\right)^2\right]+\sigma^2 \quad (1)$$

Here the expectation is the average over all possible training data sets of size $n$, which we can imagine to be randomly sampled from the total possible space of $(X_i, Y_i)$ pairs. The three right-hand side terms from left to right are the bias squared, variance, and noise variance, respectively. The bias is the difference between the expected value of our model averaged over all training set samplings $E[\hat{F}(X)]$ and the underlying true function $F(X)$. The variance is the squared spread in $\hat{F}(X)$ relative to its average, again taken over all training set samplings. Intuitively, models with few parameters that underfit but are very well constrained will minimize variance but have large bias, and models with many parameters that overfit will minimize bias but have large variance. The lowest overall errors are typically found with a balance between optimizing both the bias and the variance. Eq. (1) formally requires exploring every training data set of size $n$, and we typically have a problem with a single data set of size $n$, so it is not straightforward how to estimate the expected squared error in Eq. (1).

#### 4.4.4 Example of assessing model errors and domain of applicability using GPR and RFDT models on real data

In this section, we consider the errors and domains of some widely used modeling approaches on a realistic data set. There are two very common approaches to estimating a distribution on model prediction values. The first approach is ensemble methods, where one fits an ensemble of models, which can then yield a distribution of predictions for any new data point. The ensembles can be generated by resampling data (e.g., bootstrap and CV) or by refitting models (e.g., retraining neural networks from different starting weights or with different dropouts), or a combination of both (as is done in RFDTs), as will be described further below. The second approach is to use Bayesian methods to modify a prior distribution and produce a posterior distribution, e.g., as done in GPR. Ensemble methods are very flexible and can be applied to many models. For example, resampling can be used to get a predicted distribution for essentially any model if it is computationally feasible. Bayesian methods tend to require more specialized methods adapted to use a Bayesian approach, but can potentially avoid many iterations and include key information through priors.

In this section, we explore the behavior of error predictions from the very common approaches of CV (with GPR), GPR, and RFDTs to better see how these errors behave and might be used. For simplicity, we will usually consider just the mean and standard deviation (or RMSE, or just error) of predicted distributions, as these represent the prediction and a simple error bar, respectively, but the methods discussed here actually give a full distribution for predicted values. All these methods for estimating the error of a model result in model predicted errors on any data point. However, CV is a resampling method generally only used to predict the left-out validation samples, not totally new data, and its results are typically averaged over all predictions to obtain a single CV RMSE, as we will do here. For each case below, we illustrate the accuracy of the estimated standard deviations by comparing them to actual observed standard deviations on validation and test data sets. We will make use of models fitted to a database of DFT-calculated dilute impurity diffusion activation energies in a range of metal hosts. The data contains 408 activation energies for 15 different hosts, and is described in detail in Ref. [35] (see Sec. 7 for data availability on Figshare). All of the models were evaluated using the routines available in the scikit-learn package,[156] and the model fits and analysis were automated using the Materials Simulation Toolkit for Machine Learning (MAST-ML).[157,158]

To help assess the model domain of applicability, we explore a chemistry test where we consider Pd-X systems, where Pd is the host element and X is a dilute impurity taken from three sets (set 1 = 3d and 4d transition metals, set 2 = Col VIA elements except O, set 3 = elements from the first 2 rows on the periodic table). In this test we train the model with no Pd host data and then predict the errors for the 3 sets. While we have DFT data for only some of these predictions, they represent data that is very similar to our database (set 1, which has many 3d and 4d metals) and from quite to extremely different (sets 2 and 3, respectively), with set 2 sharing related chemistry due to being in the same column of the periodic table and set 3 having many dramatically distinct chemistries, e.g., Pd-O. Thus, we expect errors to be small in group 1, larger and similar in set 2, and larger and often outside the model domain in set 3.

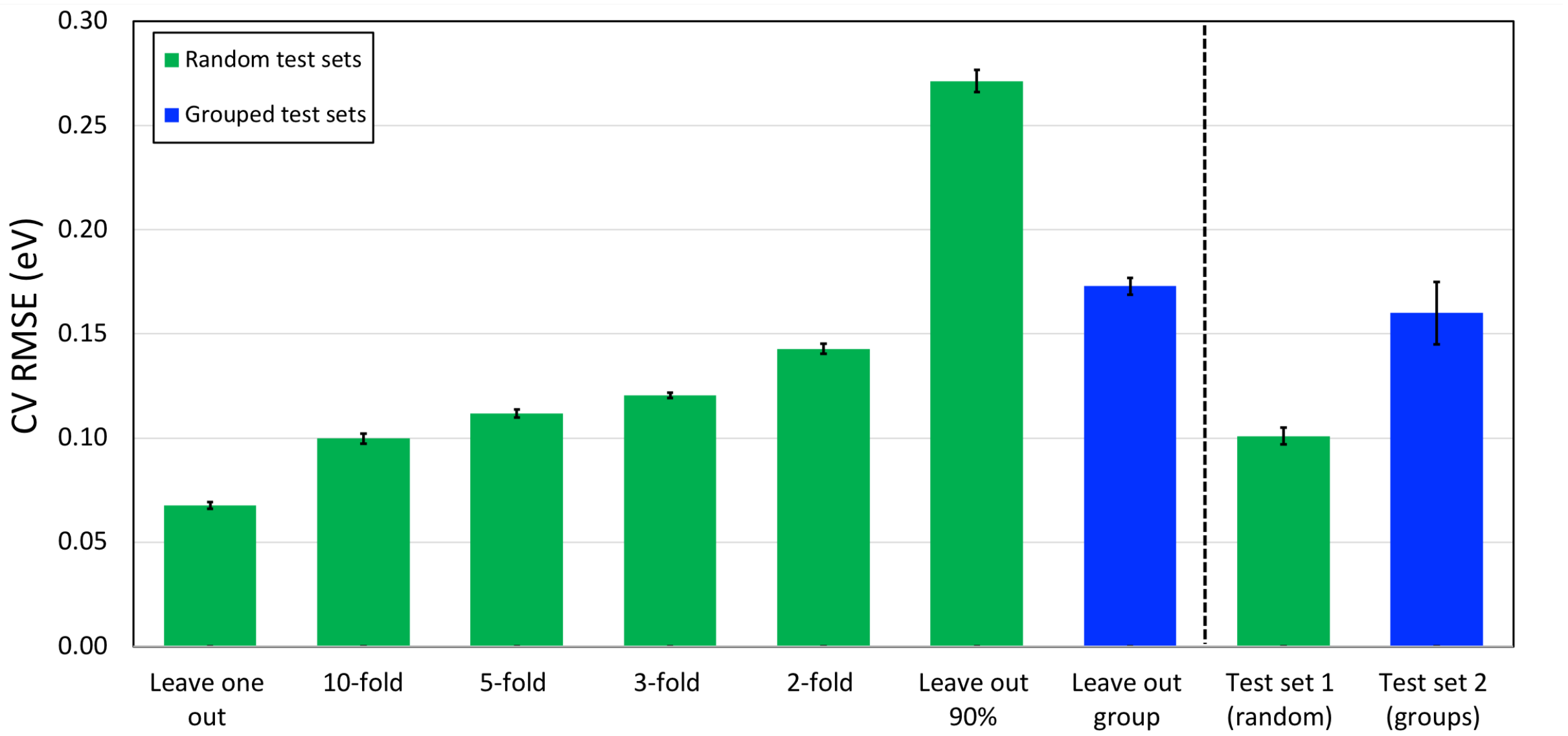

Figure 3: Plot of CV RMSE for various leave-out nested CV tests. The top level 1 split was into train/validation (left of vertical dashed line) and test data sets (right of vertical dashed line). These level 1 splits were done in two ways. Green bars signify tests done with 5 test sets chosen randomly with replacement, where each test set had 20% of the data and each train/validation set had 80% of the data. Blue bars signify tests with 15 test sets (one for each host), where each test set had one host and all impurities in that host and each train/validation set had all other data in the database. See Sec. 7 for data set availability. Predictions of test data were done with training on the full train/validation sets. The nested level 2 splits were done within the train/validation sets (left of vertical dashed line). The level 2 splits included leave one out, k-fold CV, and leave out 90% (randomly sampled 5-times, with replacement) (green bars) and leave out each host (14 splits) (blue bar). Error bars denote standard errors in the mean CV RMSE over level 2 splits (left of vertical dashed line) and level 1 splits (right of vertical dashed line). Error bars on the CV RMSE values are one standard error in the mean of the CV RMSE calculated with all values from level 2 or level 1 fits, as appropriate. All fits were done with Gaussian process regression using features optimized for this method taken from Lu et al.[35]

Perhaps the most widely used approach for estimating model errors are through the use of resampling methods, which estimate the uncertainty of predictions by sampling a subset of the available data (training data) and predicting remaining data left-out of the subset (validation data). The errors on the left-out validation data are then used to estimate a typical error bar for the model. These approaches have the advantage of being relatively simple and applicable to any model being used. The most common resampling method for error prediction is probably CV (Sec. 4.4.2). Another common resampling method is bootstrapping, which differs from CV primarily by resampling with replacement and the typical size of the resampled set. We do not discuss bootstrap in detail here due to space limitations and the fact that CV appears to have some advantages versus bootstrap for resampling.[159] However, bootstrapping is used in the random forest method described below. In addition, basic *k*-fold CV has been shown to give relatively good estimates of errors[44] and is a recommended standard test for any model. Note that for *k*-fold CV this error will generally increase with decreasing *k* (equivalently, increasing X% LO), particularly for smaller data sets, as the smaller and more independent training sets will lead to larger bias and variance. *k* in the ranges 3-10 are generally found to be a good compromise and yield good results. We illustrate this behavior in Figure 3, which shows a clearly increasing average CV error with *k* that matches the test data error best for *k* near 10.

In general, all resampling methods suffer from some significant limitations that are not always appreciated. The most severe and difficult to treat is that these methods give an estimate of the error for the data you have in your analysis (i.e. data in the training and validation sets), which error can only be expected to be accurate for data in some way similar to your database. Unfortunately, resampling does not provide a clear guide on how similar new data is to that in the database. A related issue is that when you assess an error from a LO validation data point you typically don't know how similar that point is to data in the subset used as training data. While duplicate data can be easily removed, the validation data can be very similar to one or more elements of the training data, which will typically yield errors much lower than for a prediction on a data point less similar to the validation data (this is sometimes called the twin problem, as your validation data point has one or more nearly identical twins in the training data). Both of these issues are closely related and arise from the fact that resampling yields error estimates potentially closely tied to the specific characteristics of the data sampled and predicted and may not represent the errors one will obtain for the future predictions to be made by the model. An excellent example of this problem can be found in a recent study of superconducting temperatures,[20] where models fit to just low or just high temperature superconductors both showed good cross validation scores within each group, but essentially no ability to predict the other group. This result is easy to understand in terms of the known large qualitative differences in the physics governing low and high temperature superconductivity, but one cannot rely on such robust physical guidance in general. These issues can be somewhat alleviated by careful LO group error bar assessments, where one attempts to mimic the types of prediction challenges the model will face in real applications.[14,20,35,160] For materials systems, good LO group tests might typically include leaving out certain elements, alloys, or composition ranges. For example, the LO host error on test data shown in Figure 3 is significantly larger than that obtained from the *k*-fold CV for typical *k* values of 3 or 5, demonstrating that the latter is unreliable for predicting new hosts, but it is well estimated by the LO host error determined from the training/validation data. A more direct way to avoid the twin problem might be to remove all compositions within some hypersphere around any point in the validation data, thereby ensuring the predictions are always being made from significantly different compositions. A particularly elegant way to select LO groups that mimics how your model will be used is to explicitly test new data based on data from earlier times (time-split cross validation[153]), although this is not always practical or appropriate. Sheridan used QSAR data to show that time-split cross validation was quite accurate, while random LO CV tests tended to result in an overly optimistic assessment of a model and LO clusters CV (i.e. a variant of LO group CV) tended to result in an overly pessimistic assessment of a model.[153] In general, we would recommend that all model development and error quantification done with resampling, e.g., nested CV (Sec. 4.4.2), at least use a CV error determined by combining leave out random folds and leave out physically motivated groups that assess your planned uses for the model and remove twin effects.

As mentioned above, Bayesian methods can provide an error bar without resampling. Perhaps the most widely used Bayesian method in MS&E is GPR, discussed in the ML models section in Sec. S4 of the SI. GPR distributions for a new point are entirely determined by the feature matrix of the training data and the model kernel, and do not depend on the specific values of the training data (except for a fit scale factor, $\sigma_{max}$, that typically closely matches the training data standard deviation), making GPR distributions effectively a measure of how similar a new data point to be predicted is to the database being used to train the GPR model. Data points very similar to those in the training database will have small errors, while those less similar will have larger errors. These error bars do have the limitation that they are estimates from a modified prior and are therefore expected to get less accurate for data points far from the training data. In fact, GPR error bars tend to have the constant value $\sigma_{max}$ for points completely unrelated to the original data set. Thus, for a good model and predicted errors significant less than $\sigma_{max}$ the error estimates can potentially be taken as reliable, but for predicted errors near to $\sigma_{max}$ the errors cannot be taken as quantitative, although they do suggest that the model is not robust for that data. In this way, for any prediction, GPR potentially provides either reasonable error bar estimates or a clear warning that a particular data point is outside the domain of the model. GPR error estimates can also be used to assess where the GPR model is least constrained, suggesting where a new data point might be added to best improve the model, making it a powerful guide for iterative optimization with active learning (see Sec. 3.1.2).

Figure 4A shows the standard deviations predicted for the three chemistry groups discussed above in this section, and the results are astonishingly close to what we would expect from chemical intuition. These results suggest that, at least in this case, the GPR errors are both accurate on average in the domain of the model and capable of establishing set 1 (3) as inside (outside) the model domain, with set 2 at the border of the model domain. Furthermore, Figure 4C shows, at least in this case, that the root mean of the squared residuals (RMS residuals) and GPR predicted errors show very limited correlation, suggesting that while in the model domain the GPR errors are of the correct average size, they do not appear to be varying by data point in a physically meaningful way. The results of Figure 4C suggest that GPR can predict large errors for systems well predicted by the model, so GPR may give a fairly conservative estimate for the model domain.

One of the most widely used ensemble approaches (in addition to CV) in MS&E are RFDTs, which are formulated in such a way that they provide an intrinsically powerful tool for estimating uncertainties. RFDTs train an ensemble of models and thereby predict a distribution of values for new data points, generally providing both good estimates from mean values and uncertainties from the spread of the distribution (see the ML model section in Sec. S4 of the SI for more information). The ensemble of models comprising RFDTs is traditionally generated by fitting to different data samplings (e.g., bootstrap aggregation, or bagging being perhaps the best known approach) or iteratively reweighting the fitted data to harder cases (boosting), but can also be generated from varying the model used in fitting (e.g., changing dropouts in neural networks or possible split criteria in decision trees). A detailed discussion of these approaches across all methods is outside the scope of this review but it is useful to be aware of a few important examples.

A particularly rigorous formulation of RFDT error estimates (which includes correction for the sampling and limited ensemble size as well as for missing bias and noise contributions) and an assessment showing their accuracy on materials properties is given in Ref. [161], although here we will simply use the standard deviation of the distribution of predicted values to get errors. Similar to GPR, these estimates are expected to become less accurate for data far from the original training data. For RFDTs that use the mean of the individual decision tree estimators for regression this value is bounded at half the range of the training data (since maximally varying predictions will match the lowest and highest values half the time each), although it is unlikely to reach that value and we here assume that any value approaching the standard deviation of the training data, $\sigma_{train}$, is likely to signify the data is outside the domain of the model. Unlike the GPR case, the RFDT error predictions are likely to be sensitive to both the $X$ and $Y$ values in the training data. Figure 4B shows the analogous chemistry plot for RFDTs as was shown for GPR in Figure 4A. However, unlike GPR, the RMS residuals and RFDTs predicted standard deviations show strong correlation, as shown in Figure 4D, for predicted standard deviations up to about the standard deviation of the total data set, and then show a clear transition to comparatively noisy behavior with little correlation. Also, unlike GPR, the Pd-X predictions show no ability to distinguish chemistries. These studies suggest that, for the data studied here, GPR errors are good for determining a conservative estimate for the model domain and good on average in that domain, but not reliable for distinguishing trends between data points in the domain, while RFDT errors are good on average and for distinguishing trends between data points in the domain, but not so good at estimating true errors when they approach the standard deviation of the data set and not very good at determining the model domain itself. We reiterate that these studies were done on just one fairly small data set and absolutely cannot be used to make robust broad conclusions, but the results suggest some of the opportunities and challenges of using error estimates and show the need for further studies to establish how they can be best applied to problems in MS&E.

Finally, we note that neural networks can also provide their own uncertainty estimates through an ensemble of networks approach. This can include simply starting from random weight initializations multiple times (which can be time consuming)[162], using snapshots taken during a typical optimization run[162], and exploring multiple fits done with different dropouts (dropouts in NNs are removing output of a random and changing subset of nodes).[163]

Despite one's best efforts using methods like above it can be difficult to be sure one has a meaningful model in the case of working with small data sets. A few checks against simple naïve references are recommended to ensure that the model is adding significant value. These are described in SI Sec. S5.

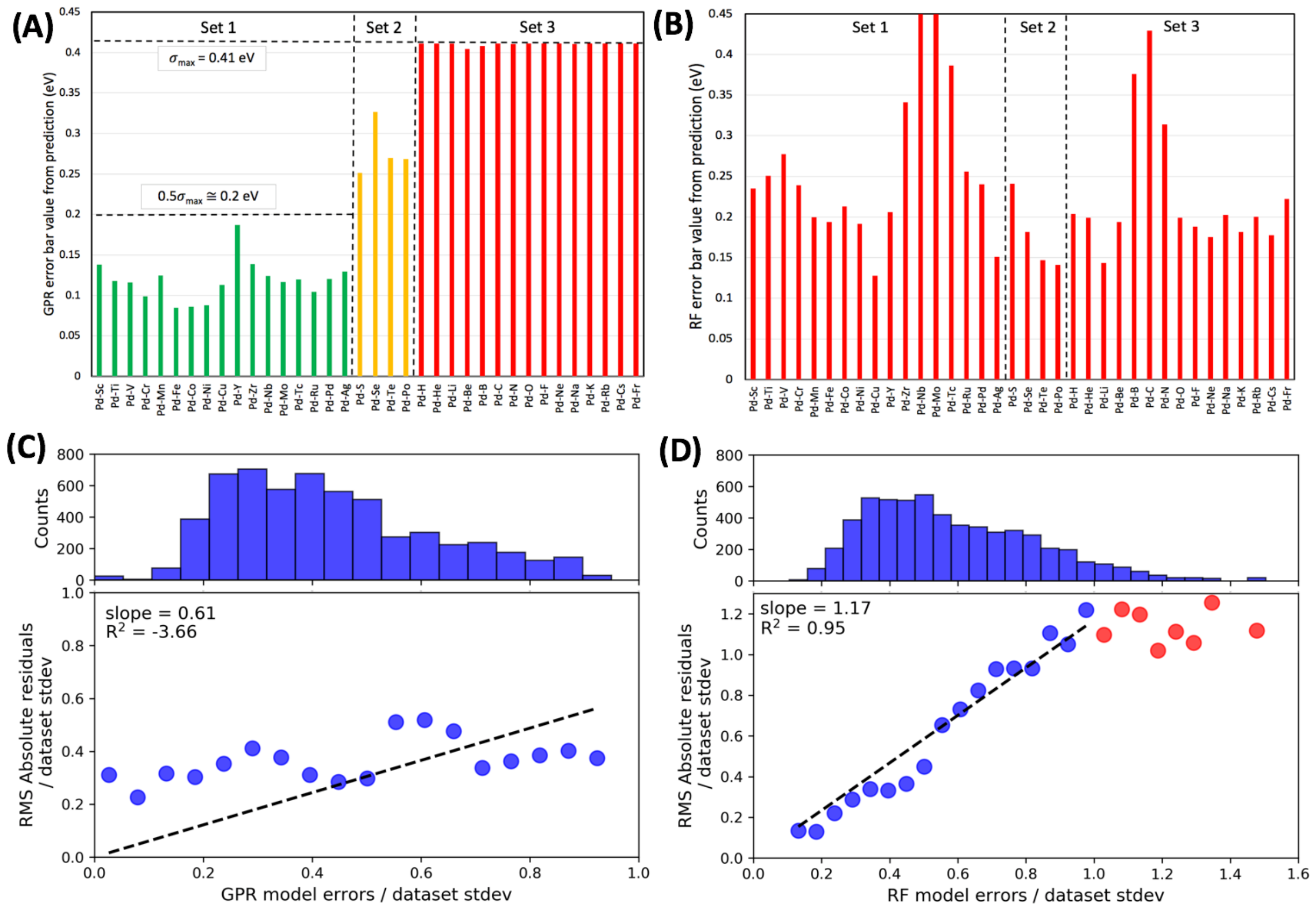

Figure 4: Summary of GPR and RFDT chemistry tests. (A) and (B) Chemistry tests showing model errors on predicted values for various solutes in a Pd host using the GPR model (A) and RFDT model (B). Comparison of root mean square absolute value of the residuals versus the binned model error values for (C) GPR and (D) RFDT tests. In (A) and (B), the models were trained on all data except Pd. In (C) and (D), both the x- and y-axes values are normalized by the dataset standard deviation, which is 0.4738 eV. The linear fits have intercepts that are forced to equal 0. In (D), the linear fit is done only on the blue data points, which have normalized binned RFDT less than 1. The histograms in (C) and (D) show the counts of the number of mean squared residuals used to obtain the RMS residual for a given model error bin. The fits in (C) and (D) were performed using the same 15 grouped datasets as described in the caption of Figure 3. These datasets are equivalent to leave out two hosts cross-validation, where each training dataset excludes two hosts and the predictions are done on the two excluded hosts. This resampling corresponds to 15×14 = 210 training/validation splits, and each data point is predicted 14 times for a total of 408×14 = 5712 total predictions. See Sec. 7 for dataset availability.

# 5 Machine learning tools and software for materials

Recently, there has been intense development of open source software packages aimed at streamlining and accelerating the adoption of ML in general, and in MS&E in particular. Effective software tools are becoming increasingly important in order to maintain community best practices and ease-of-use, especially given the rapidly evolving field of ML and its application to MS&E more specifically, and especially for users new to the field.[68,74,164] We have provided a detailed list of these packages with a brief explanation on the types of ML-related analysis enabled by each package in Sec. S2 of the SI, and posted via Figshare (see link in Sec. 7) to enable updates to this evolving list in the future.

# 6 Future opportunities and ongoing challenges of ML in MS&E

MS&E is still just beginning to utilize informatics on large databases[164,165] but the increasing data generation rates from both experiments and simulation increasingly creates opportunities, and sometimes necessitates, using ML for analysis. This trend, along with the rapid evolution of ML algorithms and supporting hardware and cloud data and computing resources, suggest that opportunities for ML in MS&E are still far from being fully realized. Here, we highlight what we see as three (of no doubt many) key opportunities and associated challenges for ML in MS&E to address in the coming years.

The first opportunity revolves around the creation of a codified, living materials informatics ecosystem which unifies materials data, MS&E-centric ML tools, and the generation, analysis, and dissemination of ML models in a democratized fashion. The development and dissemination of models in a robust innovation infrastructure is still missing, and would dramatically increase the utilization and impact of ML on MS&E. As a testament to the importance of seizing this opportunity, the potential impact and need for additional developments in ML across many fields of MS&E has been recognized in reports and at workshops hosted by many organizations, for example the Department of Energy[166], National Institute for Standards and Technology, American Society of Mechanical Engineers,[167] the National Science Foundation[167] and has been reviewed in various places.[39,68,74,164,165,168–170] As ML tools become ubiquitous in MS&E, we envision this new infrastructure would enable materials researchers, particularly the many who are not ML specialists, to construct multistep, automated workflows for complex analysis and to experiment with various algorithms and approaches to solve a particular problem, all within a consistent interface and nomenclature that implements best practices for materials-specific data, and without repeated human intervention for data formatting and translation. Such infrastructure is also necessary to allow ML models to be disseminated effectively in the broad materials innovation ecosystem, which includes ensuring they are discoverable, reproducible, reusable, and machine/human accessible, including access via an application programming interface (API) for incorporation into more complex workflows.

In addition to this new infrastructure centered on ML models, there is also a need for open-data that is curated and hosted, which will prevent data siloing and improve ease of access and sharing.[68] Consistent materials metadata, for example as implemented by the Citrination platform,[171] will also enable more informed comparisons between similar datasets, for example when comparing materials property data obtained from DFT calculations of different levels of fidelity. A long-standing challenge is related to the tradition in the scientific community where typically failed or "null" results are rarely reported in the literature. However, such results still constitute valuable information, particularly for training ML models, which can be leveraged to facilitate new materials advances. For example, recently the exploration of vanadium selenide materials synthesis was informed from failed synthesis approaches.[172] Information that is often not deemed publishable in traditional peer-reviewed scientific studies can improve ML approaches by reducing the biases toward particular outcomes of data typically reported in the literature (e.g. data on solid-state Li electrolytes may be biased toward systems that are fast Li conductors), and thus should still be made publicly available, perhaps by way of the codified infrastructure described above or through new incentives encouraged by journal publishers.

API: application programming interface
AI: artificial intelligence

In the coming years, the advancements made in the ML and broader field of artificial intelligence (AI) will likely change how humans conduct scientific research. Indeed, the advent of autonomous robot scientists has already began to shift the role of human scientists in the lab from actively conducting individual experiments to instead analyzing vast amounts of automatically produced data. These advancements create a large opportunity for more efficient and less error prone scientific investigation but will also create challenges related to how human researchers use and interact with ML/AI tools in a manner that results in improved outcomes compared to purely human- or machine-driven analysis. Human-ML collaboration (also referred to as "centaur approaches", interactive ML or "human-in-the-loop" ML) will likely evolve substantially in the near future and play a key role in many domains. For example, while computers equipped with ML/AI tools are better than humans on average for many tasks (e.g. image recognition), edge cases can still occur which result in incorrect model predictions, which cases could be quickly checked and fixed based on human intuition. Thus, human-in-the-loop ML approaches will remain useful for error minimization and sanity checks, particularly for situations where data is sparse or the edge of the domain of applicability is being reached, and will be of particular importance in situations such as the healthcare field where decisions reached using ML tools can result in life or death.[173] As a concrete example of the power of human-in-the-loop methods in MS&E, the work of Duros et al.[174] showed that active learning approaches incorporating a machine and human hybrid team outperformed both the pure human and pure ML-based prediction of performing the chemical reaction of the self-assembly and crystallization of polyoxometalate clusters. As another example, the work of Gomez-Bombarelli et al.[175] found thousands of promising organic light-emitting diode molecules in part by leveraging domain expert opinion of which molecules were most worth investigating experimentally using an online voting process. While it is currently the case that human-machine hybrid teams tend to result in better outcomes than what either humans or machines could produce in isolation, we speculate that it is very likely in the future (it is unclear when, but perhaps in the coming few decades) that ML/AI approaches will *always* outperform humans at numerous computationally intensive tasks integral to the scientific enterprise. It is also possible in the near future that how we perceive of human-in-the-loop ML may change dramatically. Instead of the human and ML algorithm being used collectively, but existing separately, it is possible that linkage of human and machine via brain-to-machine interfacing, for example as being developed by companies such as Neuralink, will fundamentally alter how human researchers interact with and use ML/AI approaches to advance the scientific enterprise.

To conclude, we see many ways in which ML (and AI) is already changing MS&E, but believe their interaction are still in the nascent stages, with the full power of their merging still far from being fully realized. The impact of their coupling is also expected to evolve quickly through building on the rapid evolution of the broader ML ecosystem, providing the opportunity for transformative advances to the discovery, design and deployment of new materials impacting myriad technologies central to today's society.

# 7 Data Availability

The diffusion activation energy dataset used in Sec. 4.4.4 is taken from the work of Lu et al.[35] and is available on Figshare (DOI: https://doi.org//10.6084/m9.figshare.7418492).

The data subsets used for training, validation and testing used in this study, data used to make each original figure, and most up-to-date supporting information document are available on Figshare (DOI: 10.6084/m9.figshare.9546305).

**Disclosure Statement**

The authors are not aware of any affiliations, memberships, funding, or financial holdings that might be perceived as affecting the objectivity of this review.

**Acknowledgments**

The authors gratefully acknowledge support of this research by NSF through the Software Infrastructure for Sustained Innovation (SI2) award no. 1148011 (which supported aspects of the machine learning tools and data used in this work), Designing Materials to Revolutionize and Engineer our Future (DMREF) award no. 1728933 (which supported aspects of the understanding of model development and fitting and relevant collaborations) and the University of Wisconsin Materials Research Science and Engineering Center (DMR-1720415) (which supported projects related to QSAR/QSPR which informed this work).

**Supporting Information**

Supporting information (SI) is available online from the publisher and on Figshare (see Sec. 7), The SI contains a list of recent ML in MS&E review papers (SI Sec. S1), detailed lists of available ML-centric software packages useful for ML in MS&E (SI Sec. S2), a short discussion of journals publishing ML in MS&E studies (SI Sec. S3), a discussion of different types of commonly used machine learning models (SI Sec. S4), a discussion on basic model checks (SI Sec. S5), and a comparison of computed residuals vs. model errors and the cumulative distribution of the ratio of residuals to model errors for the GPR and random forest model analysis presented in this paper (SI Sec S6).

Supplemental Material for
Opportunities and Challenges for Machine Learning in Materials Science
*Annual Review of Materials Research*, Vol. 50

Dane Morgan and Ryan Jacobs

Department of Materials Science and Engineering
University of Wisconsin-Madison, Madison, WI, 53706
ddmorgan@wisc.edu, rjacobs3@wisc.edu

# 8 Supplemental Section 1: Recent reviews of ML in MS&E

Given the explosion of interest and advancement of machine learning (ML) as a whole and ML in Materials Science and Engineering (MS&E) in particular, we note here that numerous reviews, progress reports, perspectives, and tutorials covering various aspects of the application of ML in MS&E have been written in just the past few years. Here we provide a list as a resource for interested readers. Note that for the purposes of this review list, we have focused on reviews specific to materials science, and thus have not made an effort to include reviews in related fields such as cheminformatics. Note that the reviews in this table are listed by year of publication, beginning with the earliest. A version of this list is also provided on Figshare (see Data Availability in Sec. 7 in the main text for link), which version can be continually updated in the future.

| Author | Publication Year | Reference | Title |
|---|---|---|---|
| Rajan | 2015 | 176 | Materials Informatics: The Materials "Gene" and Big Data |
| Broderick and Rajan | 2015 | 177 | Informatics derived materials databases for multifunctional properties |
| Kalidindi | 2015 | 178 | Data science and cyberinfrastructure: critical enablers for accelerated development of hierarchical materials |
| Mueller et al. | 2016 | 42 | Machine learning in materials science: Recent progress and emerging applications |
| Jain et al. | 2016 | 179 | New opportunities for materials informatics: Resources and data mining techniques for uncovering hidden relationships |
| Hill et al. | 2016 | 164 | Materials science with large-scale data and informatics: Unlocking new opportunities |
| Agrawal and Choudhary | 2016 | 169 | Perspective: Materials informatics and big data: Realization of the "fourth paradigm" of science in materials science |
| Le and Winkler | 2016 | 180 | Discovery and Optimization of Materials Using Evolutionary Approaches |
| Takahasi and Tanaka | 2016 | 181 | Materials informatics: a journey towards material design and synthesis |
| Kalidindi et al. | 2016 | 182 | Vision for Data and Informatics in the Future Materials Innovation Ecosystem |
| Sun et al. | 2016 | 183 | Statistics, damned statistics and nanoscience-using data science to meet the challenge of nanomaterial complexity |
| Yosipof et al. | 2016 | 184 | Materials Informatics: Statistical Modeling in Material Science |
| Audus and de Pablo | 2017 | 185 | Polymer Informatics: Opportunities and Challenges |
| Voyles | 2017 | 73 | Informatics and data science in materials microscopy |
| Liu et al. | 2017 | 170 | Materials discovery and design using machine learning |
| Ramprasad et al. | 2017 | 186 | Machine Learning and Materials Informatics: Recent Applications and Prospects |
| Ward and Wolverton | 2017 | 114 | Atomistic calculations and materials informatics : A review |
| Meredig | 2017 | 187 | Industrial materials informatics: Analyzing large-scale data to solve applied problems in R&D, manufacturing, and supply chain |
| Goh et al. | 2017 | 188 | Deep learning for computational chemistry |
| Lookman et al. | 2017 | 54 | Statistical inference and adaptive design for materials discovery |
| Lu et al. | 2017 | 189 | Data mining-aided materials discovery and optimization |
| Rupp, von Lilienfeld and Burke | 2018 | 190 | Guest Editorial: Special Topic on Data-Enabled Theoretical Chemistry |
| Correa-Baena et al. | 2018 | 191 | Accelerating Materials Development via Automation, Machine Learning, and High-Performance Computing |
| Dimiduk et al. | 2018 | 74 | Perspectives on the Impact of Machine Learning , Deep Learning , and Artificial Intelligence on Materials , Processes , and Structures Engineering |
| Tabor et al. | 2018 | 59 | Accelerating the discovery of materials for clean energy in the era of smart automation |
| Butler et al. | 2018 | 192 | Machine learning for molecular and materials science |
| Gubernatis and Lookman | 2018 | 193 | Machine learning in materials design and discovery: Examples from the present and |
| | | | suggestions for the future |
| Ye et al. | 2018 | 194 | Harnessing the Materials Project for machine-learning and accelerated discovery |
| Seko et al. | 2018 | 195 | Progress in nanoinformatics and informational materials science |
| Senderowitz and Tropsha | 2018 | 196 | Materials Informatics |
| Nash et al. | 2018 | 197 | A review of deep learning in the study of materials degradation |
| Ferguson | 2018 | 198 | Machine learning and data science in soft materials engineering |
| Sanchez-Lengeling and Aspuru-Guzik | 2018 | 69 | Inverse molecular design using machine learning: Generative models for matter engineering |
| Cao et al. | 2018 | 199 | How to optimize materials and devices via design of experiments and machine learning: Demonstration using organic photovoltaics |
| Jose and Ramakrishna | 2018 | 200 | Materials 4.0: Materials big data enabled materials discovery |
| Xu | 2018 | 201 | Accomplishment and challenge of materials database toward big data |
| Schleder et al. | 2019 | 202 | From DFT to machine learning: recent approaches to materials science–a review |
| Wan et al. | 2019 | 203 | Materials Discovery and Properties Prediction in Thermal Transport via Materials Informatics: A Mini Review |
| Rickman et al. | 2019 | 204 | Materials informatics: From the atomic-level to the continuum |
| Balachandran | 2019 | 205 | Machine learning guided design of functional materials with targeted properties |
| Gomes et al. | 2019 | 206 | Artificial intelligence for materials discovery |
| Ramakrishna, et al. | 2019 | 207 | Materials informatics |
| Agrawal and Choudhary | 2019 | 208 | Deep materials informatics: Applications of deep learning in materials science |
| Himanen et al. | 2019 | 209 | Data-driven materials science: status, challenges and perspectives |
| Reyes and Maruyama | 2019 | 210 | The machine learning revolution in materials? |
| Ong | 2019 | 211 | Accelerating materials science with high-throughput computations and machine learning |
| Venkatasubrmanian | 2019 | 212 | The promise of artificial intelligence in chemical engineering: Is it here, finally? |
| Aggour et al. | 2019 | 213 | Artificial intelligence/machine learning in manufacturing and inspection: A GE |
| | | | perspective |
| Schmidt et al. | 2019 | 130 | Recent advances and applications of machine learning in solid-state materials science |
| Arroyave and McDowell | 2019 | 67 | Systems Approaches to Materials Design: Past, Present, and Future |
| Peerless, et al. | 2019 | 214 | Soft Matter Informatics: Current Progress and Challenges |
| Gu, et al. | 2019 | 215 | Machine learning for renewable energy materials |
| Chen and Gu | 2019 | 216 | Machine learning for composite materials |
| Boyce and Uchic | 2019 | 217 | Progress toward autonomous experimental systems for alloy development |
| Barnard, et al. | 2019 | 218 | Nanoinformatics, and the big challenges for the science of small things |
| Wang et al. | 2019 | 219 | Symbolic regression in materials science |
| Dimitrov et al. | 2019 | 220 | Autonomous Molecular Design: Then and Now |
| Zhou et al. | 2019 | 221 | Information fusion for multi-source material data: Progress and challenges |
| Childs and Washburn | 2019 | 222 | Embedding domain knowledge for machine learning of complex material systems |
| Ceriotti | 2019 | 223 | Unsupervised machine learning in atomistic simulations, between predictions and understanding |
| Lamoureux et al. | 2019 | 224 | Machine Learning for Computational Heterogeneous Catalysis |
| Lookman et al. | 2019 | 225 | Active learning in materials science with emphasis on adaptive sampling using uncertainties for targeted design |
| Chan et al. | 2019 | 108 | Machine Learning Classical Interatomic Potentials for Molecular Dynamics from First-Principles Training Data |
| Hase et al. | 2019 | 61 | Next-Generation Experimentation with Self-Driving Laboratories |
| McCoy and Auret | 2019 | 226 | Machine learning applications in minerals processing: A review |
| Faber and von Lilienfeld | 2019 | 227 | Modeling Materials Quantum Properties with Machine Learning |
| Ball | 2019 | 228 | Using artificial intelligence to accelerate materials development |
| Vasudevan, R., et al. | 2019 | 229 | Materials science in the artificial intelligence age: high-throughput library generation, machine learning, and a pathway from correlations to the underpinning physics |
| Wei, J., et al. | 2019 | 230 | Machine learning in materials science |
| Zhou et al. | 2019 | 231 | Big Data Creates New Opportunities for Materials Research: A Review on Methods and Applications of Machine Learning for Materials Design |
| Ju and Shiomi | 2019 | 232 | Materials Informatics for Heat Transfer: Recent Progresses and Perspectives |
| Jackson et al. | 2019 | 233 | Recent advances in machine learning towards multiscale soft materials design |
| Bock et al. | 2019 | 234 | A review of the application of machine learning and data mining approaches in continuum materials mechanics |

# 9 Supplemental Section 2: Software tools to enable and enhance ML in MS&E

ML: machine learning
MS&E: materials science and engineering

Recently, there has been intense development of open source software packages in ML, and more specifically those aimed at streamlining and accelerating the adoption of materials informatics research. Software tools, especially given the rapidly evolving field of ML and its application to MS&E more specifically, are becoming increasingly important in order to maintain community best practices and ease-of-use, especially for users new to the field.[68,74,164] An extensive list of software packages are listed in this section, along with a brief explanation on the types of ML-related analysis enabled by each package. We note that this list is not comprehensive and new packages appear frequently, but we believe the list should be useful for those trying to make sure they are aware of available tools. Overall, the type of software package may be categorized into one of eight groups, where group 1 denotes are ML environments with many packages pre-installed, groups 2 and 3 denote general (multidisciplinary) software and the remaining groups denote software that is tied more specifically to ML problems in MS&E.

(1) *ML-exploration and hosting environments:*

- Google Colab: Free cloud based free Jupyter notebook environment with many ML packages preinstalled and free computer resources available.[235]
- NVIDIA NGC: Portal for a wide-range of free ML software prepared in containers for rapid GPU deployment.[236]
- Nanohub: A science and engineering API with many community-contributed resources, including ML-centric tools.[237]
- DLHub: Online center for hosting, sharing, and publication of data and ML models through a user-friendly interface.[238]

(2) *Paid commercial ML-centric software services:*

- Datarobot: Enterprise ML software enabling easy automation of entire ML analysis pipeline.[239]
- Amazon Sagemaker: Part of Amazon Web Services, Amazon's Sagemaker provides enterprise software to quickly build, train, and deploy ML models.[240]

- Microsoft Azure: The machine learning studio within Microsoft Azure contains a fully managed cloud service providing enterprise machine learning software to quickly build, train, and deploy ML models.[241]
- IBM Watson: Enterprise ML software web API.[5,242]

(3) *Open source software enabling the use of ML algorithms:*

- Scikit-learn: A Python package with a wide array of algorithms encompassing every portion of the ML analysis pipeline.[156]
- Waikato Environment for Knowledge Analysis (WEKA): A Java package with a wide array of algorithms encompassing every portion of the ML analysis pipeline.[243]
- R: A general data science and machine learning package.[244]
- TensorFlow: Designed to enable custom, complex, highly flexible neural network models.[245]
- Keras: A user-friendly front end API for TensorFlow.[246]
- PyTorch: A package enabling more widespread use of deep learning, particularly for image analysis.[247]
- ChainerCV: A library for deep learning with a focus on computer vision.[248]
- DeepChem: A library for deep learning with a focus on analysis of chemical characteristics of molecules.[143]

(4) *Software consisting of trained models enabling easy prediction of materials properties*:

- AFLOW-ML: Web-hosted ML models with drag-and-drop prediction of numerous properties.[249]
- ElemNet: A deep learning neural network trained using only elemental compositions enabling the prediction of material formation energies.[250]
- JARVIS-ML: Web-hosted ML models with drag-and-drop prediction of numerous properties.[251]
- PhysNet: A deep neural network enabling predictions of energies, forces and dipole moments for small molecules.[252]

(5) *Software enabling improved feature engineering for more robust ML model generation*:

- Materials Agnostic Platform for Informatics and Exploration (MAGPIE): methods of feature generation using elemental properties.[253]
- Materials Simulation Toolkit for Machine Learning (MAST-ML): automation of ML pipeline and codifying of best practices of ML in MS&E, including data cleaning, feature engineering, model fitting, cross-validation and assessment of many statistics.[254]
- matminer: codified set of useful data visualization and structure- and chemistry-based feature generation schemes.[139]
- DScribe: codified set of structure- and chemistry-based feature generation schemes.[255]

(6) *Software streamlining ML analysis methods and the ML pipeline*

- LoLo: Automated ML model fitting and data analysis, estimates of errors based on random forest models.[171]

- matminer: (see above)
- MAST-ML: (see above)
- MATcalo: materials knowledge-based assistive software to aid researchers in MS&E using ML to conduct improved materials research.[256]
- MatErials Graph Network (MEGnet): automated construction and evaluation of graph-based convolutional neural networks for molecules and crystals.[142]
- SchNet: automated construction and evaluation of deep tensor neural networks for prediction of molecule and crystal properties.[148]
- Veidt: streamlined construction of deep learning neural networks for materials science.[257]
- Materials Knowledge Systems in Python project (pyMKS): ML analysis of structure-property-processing relationships with a focus on microstructure characterization.[258]
- Tree-based Pipeline Optimization Tool (TPOT): automation of ML pipeline, particularly choice of best ML model.[259]

(7) *Software facilitating the creation of interatomic potentials*

- Atomic Energy Network Package (aenet): fitting neural network-based models for interatomic potentials.[260]
- Atomistic Machine Learning Package (AMP): construction of MLPs using a variety of atomic structure descriptors and machine learning models.[261]
- SimpleNN: fitting neural-network-based models for interatomic potentials.[262]
- PES-Learn: automated production of neural-network or Gaussian process models for constructing interatomic potentials.[263]
- DeePMD-kit: construction of MLPs using deep learning neural networks.[264]
- Convolutional Neural Networks for Atomistic Systems (CNNAS): creation of deep convolutional neural networks for interatomic potentials.[265]
- TensorMol-0.1: creation of interatomic potentials consisting of trained neural network combined with screened long-range electrostatic and van der Waals physics.[266]
- SchNetPack: extends SchNet and aids in creation of machine learning potentials using deep learning neural networks (using PyTorch).[149]
- sGDML: python package for force-field generation using the symmetric gradient domain machine learning (sGDML) model.[267]

(8) *Software facilitating the use of natural language processing*

- Elsevier API: Application Programming Interface (API) to Elsevier published texts to support text and data mining.[268]
- word2vec: NLP methods to efficiently construct word embeddings (that map words to real valued vectors).[79]
- Global Vector (GloVe): NLP software that combines global matrix factorization and local context window methods.[80]
- Materials science embeddings: Word embeddings trained for materials science.[87,88]
- Character to Sentence Convolutional Neural Network (CharSCNN): Tools to conduct sentiment analysis using deep convolutional neural networks.[81]

# 10 Supplemental Section 3: Journals publishing ML in MS&E studies

As ML in MS&E has greatly expanded in scope in the past several years, there are many journals that have or might be expected to publish ML-related studies in MS&E. In particular, some journals seem particularly well represented in this area and additionally appear interested in publishing papers with a relatively more methodological focus that might contain limited new materials insights. These journals include, but are not necessarily limited to (in alphabetical order): Computational Materials Science (Elsevier), Computer Physics Communications (Elsevier), Integrating Materials and Manufacturing Innovation (Springer), Journal of Chemical Theory and Computation (ACS), Machine Learning: Science and Technology (IOP), Materials Today Advances (Elsevier), Molecular Systems Design & Engineering (RSC), MRS Communications (MRS), and npj Computational Materials (Nature, open access). There are also a number of more chemistry-oriented journals that publish papers in ML and the areas of QSAR/QSPR, e.g., Chemometrics and Intelligent Laboratory Systems (Elsevier), Journal of Computational Chemistry (Wiley), Journal of Chemical Information and Modeling (ACS), Journal of Computer-Aided Molecular Design (Springer), and Molecular Informatics (Wiley). For papers where the materials insights are significant any materials journal could, of course, be appropriate.

# 11 Supplemental Section 4: Types of Machine Learning Models

This section primarily describes some standard ML models (with a focus on supervised regression) widely used in MS&E with a goal of highlighting the most salient features for an MS&E researcher. The discussion touches on basic aspects which are covered in many textbooks and general reviews, e.g., Refs. [42–44,152], and we therefore do not provide additional references unless addressing a specific feature outside the scope of these broad ML texts. We will use notation similar to the main text in that we assume our data has the original form (*X*,*Y*), where *X* is a matrix of features where each row corresponds to a system to be predicted and each element in that row is a value describing some feature of the system, and *Y* is a vector of target properties to be modeled. The relationship between *X* and *Y* can be written as $Y = F(X) + \epsilon$, where $\epsilon$ is a noise term (with mean zero and variance $\sigma^2$) and we seek a model for $F(X)$ from ML. We write this model as $\hat{F}(X)$ and its predictions as $\hat{Y}$.

## 11.1 Multivariant Linear Regression (MVLR)

MVLR assumes that the target *Y* is a linear function of the features *X*. Note that these are generally used with some kind of regularization which penalizes large variations in the fitted coefficients, either the *L2* or *L1* norm, an approach known as ridge regression. These methods are notable for being extremely fast, deterministic, and very easy to interpret (e.g., the coefficient of each term gives its effect, and magnitudes effectively rank the importance of each variable). MVLR is also simple enough that an enormous body of statistical data on the fit can be determined essentially analytically. For example, uncertainties in all fitted coefficients and their covariance, and uncertainties in any predictions, can be readily obtained, and these can include the influence of uncertainties in the data being fit. Accurate fitting with MVLR does require that the *Y* values be an approximately linear function of the features *X*, but since *X* can include arbitrary functions of underlying descriptors (e.g., polynomials, logarithms, etc.) MVLR does not require linearity with an initial set of features. For a given set of features MVLR does not have

MVLR: multivariate linear regression
KRR: kernel ridge regression
GPR: gaussian process regression
CV: cross validation
DT: decision tree
RFDT: random forest decision tree

any hyperparameters (one is introduced by regularization) although the feature engineering to introduce nonlinearity can effectively add many adjustable parameters. Because of the powerful statistical tools and extremely rapid and robust fitting enabled by MVLR, it is often desirable to consider such models first as they provide a useful baseline. However, materials properties are generally not expected to behave as a linear function of simple features or simple closed form functions of features, which means that MVLR models tend to either poorly represent the data or overfit it, leading to inaccurate estimates of values and/or large estimated uncertainties in predictions, and serious errors for data even slightly different from the data set. For these reasons MVLR is typically not the method of choice for most MS&E ML problems.

## 11.2 Kernel Methods and Kernel Ridge Regression (KRR)

Kernels are an inner product between feature vectors, which effectively define a distance between any two data points in terms of their feature vectors. This distance supports a nonlinear modeling of the data and can be used as basic input to a wide variety of ML methods, including support vector machines, principal component analysis, and spectral clustering. One of the models widely used for simple regression is KRR, which effectively predicts a new target output ($\hat{Y}^*$) from the new input features $X^*$ in terms of a linear combination of training data features $X_i$ weighted by their kernel-derived distances from $X^*$, and includes ridge regulation of the coefficients. This method is fairly fast to fit and often provides a good nonlinear model of $Y = F(X) + \epsilon$. The kernel typically introduces at least one hyperparameter. For example, the commonly used Gaussian kernel has a length scale that sets the range over which the distance metric decays, a value that must be similar to length scales within the problem feature set to obtain a good model. Because all kernels go to zero for widely separated points, KRR predicts a value of zero for all points very far from the training data (this can be shifted to predict the mean of $Y$ by normalizing $Y$ to mean of zero before fitting). It should be noted that even with just two hyperparameters for kernel length scale and regularization, one can get strong coupling between them and get families of models where similar CV performance is obtained for a wide subspace of values where the two hyperparameters are linearly correlated.[269]

## 11.3 Gaussian Process Regression (GPR)

GPR is a Bayesian approach that assumes a prior multivariate normal distribution for $Y$ values with a covariance between $Y_i$ and $Y_j$ given by the distance between feature vectors $X_i$ and $X_j$, and then modifies this distribution using the training data and Bayes theorem.[270] The distance is determined by a kernel, described in Sec. S11.2, making its predictions similar to KRR. However, GPR predicts a distribution of $\hat{Y}^*$ for any new $X^*$, and the first and second moment of the distribution can be used to estimate the predicted value and its variance (see Sec. 4.4.4 of the main text for some assessment of GPR predicted standard deviations).

## 11.4 Random Forest Decision Trees (RFDTs)

RFDTs are often the preferred method to for ML modeling for simple regression problems as they are highly accurate, very fast to train and evaluate, effectively perform their own feature selection and yield features ordered by importance, and provide intrinsic error estimates on predictions. A single DT is created by iterative splitting the data on features (nodes) so as to maximize some score metric (e.g., entropy reduction) until reaching a the end of the tree (leaf), and a tree classifies any input into a leaf. Mean values or linear fits to data within each leaf provide regression estimates. Single trees are prone to overfitting, a problem solved by the random forest approach,

NN: neural network
DNN: deep neural network
GPU: graphics processing unit

which creates an ensemble of DTs through training many DTs on partial samplings of the data (using bootstrap aggregating, or bagging) while simultaneously altering the available split criteria at nodes.[155] RFDTs therefore predict a distribution of values, one from each DT, for any new data point, and the first and second moments of this distribution can be used to predict new values and their variance. RFDTs have a number of hyperparameters (e.g. maximum depth of tree) but sensible defaults can often be chosen such that results depend only weakly on the hyperparameters. The accuracy of RFDTs often approaches that of a highly trained NNs but with a fraction of the time taken in training and hyperparameter optimization. As an ensemble method, RFDTs produce a distribution of predictions, and the first and second moment of the distribution can be used to estimate the predicted value and its variance (see Sec. 4.4.4 of the main text for some assessment of RFDT predicted standard deviations).

## 11.5 Basic Neural Networks (NNs)

NNs are in many ways the most powerful and versatile ML tools. A node takes input data, weights it, and then passes that weight through an activation function, yielding an output value. A layer has many nodes, and multiple layers can be connected. We use "basic" neural networks to refer ones with up to just a few layers that have no special processing to enable effective training of many layers or feature reduction through convolution or pooling. Networks with these additional features are called deep NNs and discussed in Sec. S11.6. Basic NNs have many adjustable weights and are typically trained by simple steepest descent optimizations, which typically yield different final weights for different weight initializations. This is in contrast to the MVLR, KRR, and GPR discussed above, which are essentially uniquely determined in a fit. NNs also have a large space of hyperparameters, like number of nodes and layers and activation function type, which can significantly impact their results. For these reasons, training an optimized robust NN is generally significantly more challenging and time consuming than any of the above methods and often yields only modest improvements. These methods are therefore typically tried after those discussed above if needed for standard MS&E regression problems.

## 11.6 Deep Neural Networks (DNNs) and Deep Learning

DNNs represent a significant step in ML that we briefly summarize here.[152,208] DNNs can be most simply thought of as a NN with many layers (hence the terminology "deep"), but to make the models effective and trainable new types of layers (e.g., convolution, pooling) and optimizations (e.g., residual fitting[145]) are used. DNNs have a number of distinct features compared to traditional methods which we briefly summarize here. Below we will frequently compare to the human brain as this makes a helpful analogy, but we do not mean to imply that these DNNs are actually working by mechanisms equivalent to our brain or make any it suggestions that they are human in some meaningful way.

1. Dimensional reduction and feature map development: Deep NNs typically involve stages that reduce the complexity of the input data, e.g., through convolutions or pooling. These allow extremely large and complex initial feature sets to be used, including ones where related data (e.g., nearby image pixels or connected atom and bond properties) are not co-located in the feature vector, and can effectively extract a reduced set of essential features without human intervention. Compared to human feature generation this process can be much faster, much easier to apply to new situations, and more accurate. Just as your brain can identify key features of a picture of a cat without defining them explicitly, so can a DNN extract the features without having to write them down in advance.

2. Highly flexible weights: Deep NNs can easily have millions of adjustable weights which gives and incredibly rich model for connecting inputs and outputs. This richness means that model fits are not unique (i.e., two users fitting to the same data with the same tools will not get the same weights). However, it also means that weights can be tuned to perform many tasks, e.g., to identify multiple properties from an input molecular structure. These weights also mean DNNs require extensive training, often done on GPUs and taking multiple hours for typical materials problems (e.g., with hundreds to thousands of data points or images).
3. Scalable fitting: The weight fitting typically uses backpropagation to push weights in the direction that minimizes a loss function (e.g. RMSE). However, such pushes can be done sequentially, allowing training on subsets of the data, generally called batches. In this way one can easily train on almost any size data by simply breaking it up into manageable batches. Our brains work similarly, learning more about how to identify a cat with each cat we see, but not needing to see all cats at once.
4. Transfer learning: The weights contain such a rich map of key features that they can often be very effectively transferred from one problem to another, a method called transfer learning. Transfer learning can reduce the size of data set needed for training in images from tens of thousands to just hundreds.
5. Data hungry: DNNs typically require large data sets to fit the large sets of weights, although transfer learning can greatly reduce these requirements.
6. Flexible architecture: DNNs are highly flexible and come in many forms. Some are just different types of layers in different orders with different connections between them, e.g., varying numbers of convolution layers or connection their output across many layers to avoid fitting problems. However, many DNNs have very profound changes compared to a simple multilayered NN. A common approach is to have multiple NN active in a single method. The faster regional convolutional neural network (Faster-RCNN) uses this approach identify objects in images, having one NN trained to propose bounding box regions for objects and another to fit the object location. Other important architectures distinct from simple layered networks include Generative Adversarial Networks (GANs) and recurrent neural networks (in particular, long-short term memory NNs), where the latter are extremely successful for data that comes in a series, e.g., time series or games or text.

> RMSE: root mean squared error
>
> RCNN- regional convolutional neural network
>
> GAN: generative adversarial network

7. Generative ability: An exciting area of DNNs for materials are generative models, which learn to propose new members of a distribution of samples and can therefore actually propose new materials no human has considered. A recent notable development in this area are Generative Adversarial Networks (GANs). GANs contain two NNs, one that proposes candidates (generator) and one which screens for real candidates (discriminator), and by training them together GANs find an optimal joint performance that can generate new examples of a class. These have been able to generate extremely realistic images of desired types as well new molecules with desired properties.[69271]

## 11.7 Methods for small datasets

Datasets in MS&E are often small, and techniques adapted to this type of data are particularly useful. a simple approach to obtain useful predictions from small datasets is to simplify the physics of the target quantity by subtracting a relevant reference, then fitting a model to this shifted target quantity.[272] A more complex but powerful approach is transfer learning, which uses results from ML on a different data set to inform the target modeling effort.[273] Transfer learning can be done by using ML to create improved features that might not be readily available, or predicting a useful reference to shift the target quantity. One can also train the same model on multiple data sets (either sequentially or simultaneously), an approach widely used in text mining and machine vision (see Sec. S11.6. In materials, pretrained DNNs have been used in a number of microstructure image processing tasks.[27475]

# 12 Supplemental Section 5: Some checks for model value against a naive model reference

In any modeling exercise it is useful at the end to check against some simple baseline reference cases to be sure the model has value. Here are a few such tests that are recommended. Note that we follow the notation introduced in Sec. S11 and Sec. 2 of the main text.

- Permuted data: We note a simple way to check if overfitting is playing a major role in the model is to permute the $Y$ values randomly so that they have no physical connection with the $X$ features (but still have identical properties in terms of sizes, distribution, etc.) and repeat the model development strategy (e.g., as done in Ref. [36]). One should obtain significantly worse performance than the unpermuted model and ideally RMSE/σ ≈ 1 and $R^2 \approx 0$.
- Dummy regressor/classifiers: Perhaps the simplest prediction model is to guess a simple value derived from the data (e.g., mean, median, constant, specific quantile). One's model should do much better than this method on all basic metrics (e.g., RMSE, $R^2$, etc.). Scikit-learn[156] implements a number of dummy classifier and regression functions.
- Nearest neighbor: A simple model for predicting $Y^*$ from $X^*$ is to take the $Y_i$ value from the $X_i$ that is closest to $X^*$.[20] Closeness can be measured by a simple Euclidean distance or some more complex kernel. This type of model may actually work quite well in some cases and it is not necessarily a problem for a develop ML model if it has a similar performance in some aspects, but it is worth being aware how accurate one can be with such a simple approach so one does not use a much more complex model unnecessarily.

# 13 Supplemental Section 6: Further comparison of assessing model errors using GPR and RFDT models on real data

For additional assessment of model errors, we will make use of the distribution of a statistical variable $r$ (called the *r-distribution*), where

$$r_i = \frac{(residual\ of\ data\ point\ i)}{(estimated\ standard\ deviation\ of\ data\ point\ \ i)} \quad \text{(S1)}$$

This variable should follow a normal distribution if the estimated standard deviations are accurate and the residuals are normally distributed.[16135] We will compare the distribution of $r$ for validation data to the normal distribution. We will represent the distributions by their cumulative values i.e., plotting the fraction of the magnitudes of $r$ (and a reference normal distribution) that are less than a given value, written formally as $x/\sigma$.

Figure S1A shows the cumulative $r$ distribution (Eq. (S1)) for the LO group test using GPR compared to that for a normal distribution. Only errors $\leq 2/3 \times \sigma_{train}$ (where $\sigma_{train}$ is the standard deviation of the training data) are included in the $r$ distribution to avoid values likely to be outside the model domain. The agreement between both curves is fairly good, supporting the qualitative accuracy of the GPR standard deviation estimates. However, the $r$-distribution does show that the standard error values are too large (small) up to (after) about 1.75 times the standard deviation for the test data. This effect leads to too many rare events, e.g., about 2% of the errors are more than three times the GPR predicted standard deviation, which is much larger than the 0.3% obtained for an analytical normal distribution. Figure S1B shows the analogous cumulative $r$-distribution (with only errors $\leq 2/3 \times \sigma_{train}$ included in the distribution) this time for RFDTs. Again, the $r$-distribution matches that from the analytical normal distribution reasonably well, although generally the error bars from the RFDT appear to be a little too large on average for $r <$ 2.

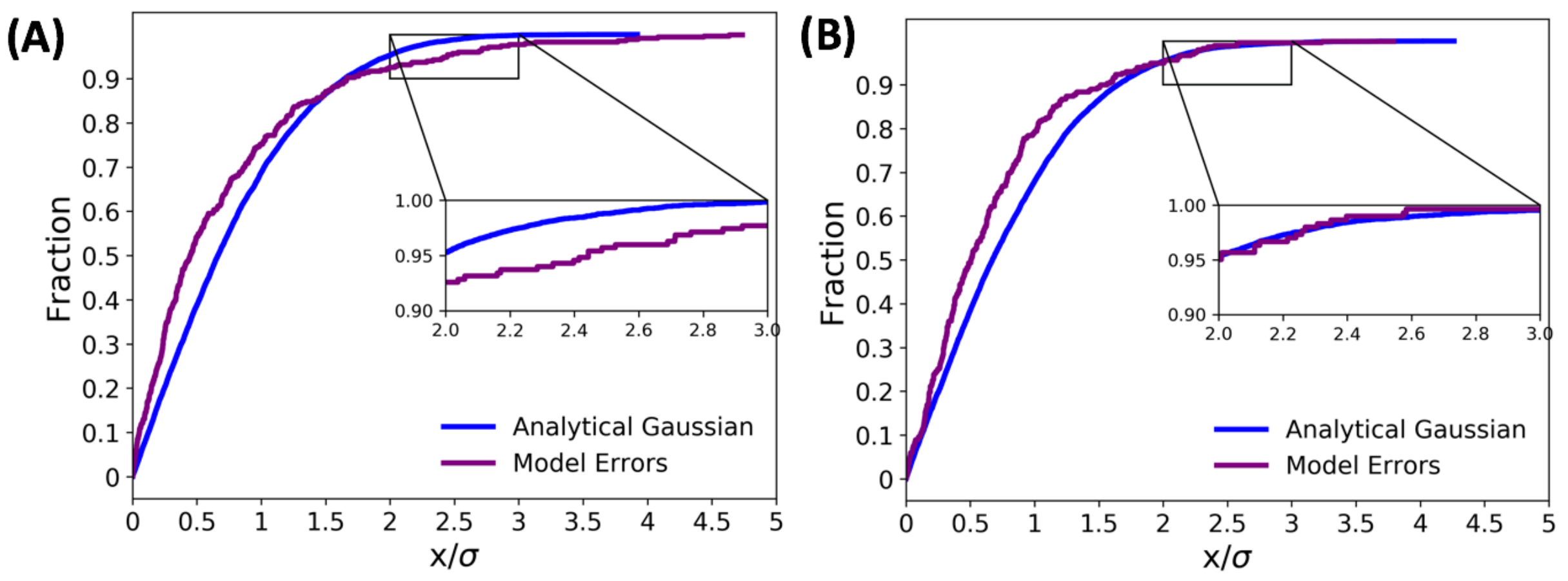

Figure S1: Cumulative $r$-statistic plots for LO group tests using GPR (A) and RFDT (B) using grouped datasets.

Figure S2 contains plots of all the absolute residuals versus the GPR (A) and RFDT (B) model errors, where both the residuals and the model error values have been normalized by the standard deviation of the dataset. The data plotted in Figure S2 are the same data used to create the RMS average "binned" plot of residuals vs. model errors (Figure 4C and Figure 4D) of the main text.

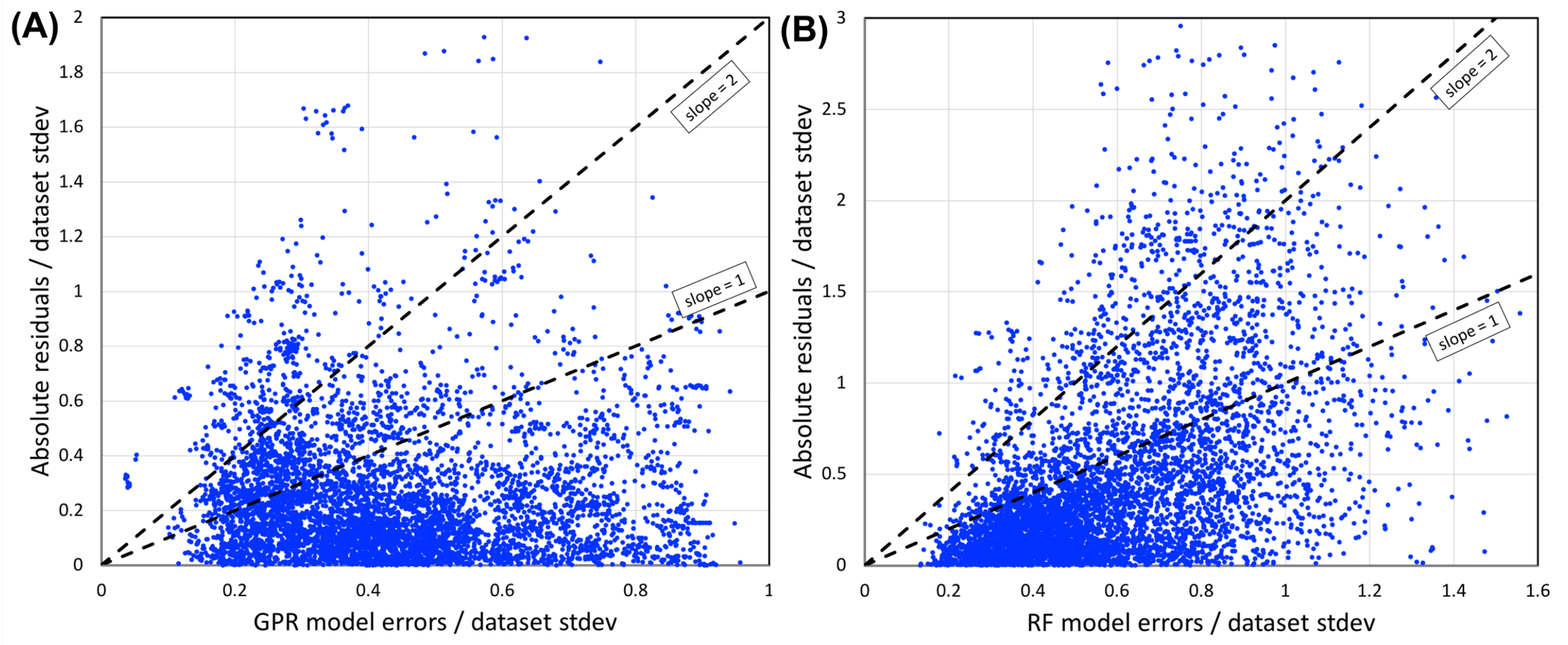

Figure S2: Comparison of absolute residuals versus the corresponding model error values for (A) GPR and (B) random forest tests. Both the x- and y-axes values are normalized by the dataset standard deviation, which is 0.4738 eV. The values are taken over leave out group tests on the same grouped datasets used in the analysis presented in the main text. The dashed lines are not fits to the data, but instead denote slopes of 1 and 2 (with a forced intercept of 0), which may be helpful to flag potential points which may be outside the domain of the model.